\documentclass[11pt]{article}

\usepackage{amscd}
\usepackage{amsmath}
\begin{document}

\newcommand{\N}{N\raise.7ex\hbox{\underline{$\circ $}}$\;$}

\title{ V.M. Red'kov, E.M. Ovsiyuk\footnote{redkov@dragon.bas-net.by; e.ovsiyuk@mail.ru} \\
On exact solutions for quantum particles\\  with spin $S= 0, 1/2,
1 $ and       de Sitter event horizon}

\maketitle

\begin{abstract}

Exact wave solutions for particles with spin $0, 1/2$ and $1$ in
 the static coordinates of the de Sitter space-time model
 are examined in detail.

Firstly, for a scalar particle, two pairs of linearly independent
solutions are specified explicitly:  running and standing waves.
 A known
 algorithm for calculation of the reflection coefficient
$R_{\epsilon j}$ on the background of the~de Sitter space-time
model is analyzed. It is shown that the
determination of  $R_{\epsilon j}$ requires an additional
constrain on quantum numbers $\epsilon \rho / \hbar c >> j$, where
$\rho$ is a curvature radius. When taken into account of this condition,
the  $R_{\epsilon j}$ vanishes identically.
 It is claimed
that the calculation of the reflection coefficient $R_{\epsilon j}$
is not required at all because there is no barrier in an~effective
potential curve  on the background of the~de Sitter space-time.

The same conclusion holds   for arbitrary particles with higher
spins, it is demonstrated explicitly with the help of exact
solutions for  electromagnetic and  Dirac fields.

%{\bf keywords:}Fields \and curved space \and  de Sitter space \and
%Dirac equation \and complex formalism by  Majorana and Oppenheimer
%\and exact solutions \and de Sitter horizon}

% \subclass{33C05 \and 34B05  \and 83C50 }
\end{abstract}

\section{ Introduction}

Examination of  fundamental  particle fields on the background of
expanding universe, in particular de Sitter and anti de Sitter
models, has a long history; special role of these  geometries
consists in their simplicity and high symmetry groups that allow
us to believe in existence of exact analytical treatment for  some
fundamental problems of classical and quantum field theory  in
these curved spaces. In particular, there exist special
representations for fundamental wave equations, Dirac's and
Maxwell's ones, which are explicitly invariant under the symmetry
groups $SO(4.1)$  and $SO(3.2)$ for these models. In  most of the
literature, when dealing with a spin 1 field  in de Sitter models,
the  group theory approach is used. Many  important references are
given below (of course, it is not an exhaustive bibliography list):
Dirac \cite{Dirac-1935}, \cite{Dirac-1936},
 Schr\"{o}dinger
\cite{Schrodinger-1939}, \cite{Schrodinger-1940},
Lubanski--Rosenfeld \cite{Lubanski-Rosenfeld-1942},
 Goto \cite{Goto-1951},
 Ikeda \cite{Ikeda-1953},
 Nachtmann \cite{Nachtmann},
 Chernikov--Tagirov \cite{Chernikov-Tagirov},
Geheniau--Schomblond \cite{Geheniau-Schomblond},
 Bor\-ner--Durr \cite{Borner-Durr-1969},
 Tu\-gov   \cite{Tugov-1969},
Fushchych--Krivsky  \cite{Fushchych-Krivsky-1969},
Chevalier \cite{Chevalier-1970},
Castag\-nino  \cite{Castagnino-1970}, \cite{Castagnino-1972},
Vidal  \cite{Vidal-1970},
Adler \cite{Adler-1972},
Schnirman--Oliveira \cite{Schnirman-Oliveira-1972},
Tagirov   \cite{Tagirov-1973},
Riordan   \cite{Riordan-1974},
 Pestov--Cher\-nikov--Shavoxina  \cite{Pestov-Chernikov-Shavoxina-1975},
 Candelas--Raine \cite{Candelas-Raine-1975},
  Schom\-blond--Spindel \cite{Schomblond-Spindel-1976}, \cite{Schomblond-Spindel-1976'},
Dowker--Critchley    \cite{Dowker-Critchley-1976},
Avis-Isham--Storey   \cite{Avis-Isham-Storey-1978},
 Brugarino \cite{Brugarino-1980},
 Fang-Fronsdal   \cite{Fang-Fronsdal-1980},
Angelopoulos et al \cite{Angelopoulos-Flato-Fronsdal-Sternheimer-1981},
Burges \cite{Burges-1984},
Deser-Nepo\-mechie \cite{Deser-Nepomechie-1984},
 Dullemond-Beveren \cite{Dullemond-Beveren-1985},
 Gazeau  \cite{Gazeau-1985},
 Allen \cite{Allen-1985},
 Fef\-ferman--Graham \cite{Fefferman-Graham-1985},
 Flato--Fronsdal--Gazeau
\cite{Flato-Fronsdal-Gazeau-1986}, Allen--Jacobson
\cite{Allen-Jacobson-1986}, Al\-len--Folac\-ci
\cite{Allen-Folacci-1987}, Sanchez    \cite{Sanchez-1987},
Pathinayake--Vilenkin--Allen
\cite{Pathinayake-Vilenkin-Allen-1988}, Gazeau--Hans
\cite{Gazeau-Hans-1988}, Bros--Gazeau--Mo\-schella
\cite{Bros-Gazeau-Moschella-1994}, Takook  \cite{Takook-1997},
Pol'shin  \cite{Pol'shin-1998(1)}, \cite{Pol'shin-1998(2)}, \cite{Pol'shin-1998(3)}, Ga\-zeau--Takook  \cite{Gazeau-Takook-2000},
Takook   \cite{Takook-2000}, Deser--Waldron
\cite{Deser-Waldron-2001}, \cite{Deser-Waldron-2001'},
Spradlin--Strominger--Volovich
\cite{Spradlin-Strominger-Volovich-2001}, Cai--Myung--Zhang
\cite{Cai-Myung-Zhang-2002}, Garidi--Huguet--Renaud
\cite{Garidi-Huguet-Renaud-2003}, Rouhani--Takook
\cite{Rouhani-Takook-2005}, Behroozi et al
\cite{Behroozi-Rouhani-Takook-Tanhayi-2006}, Huguet--Queva--Renaud
\cite{Huguet-Queva-Renaud-2006},
 Garidi et al   \cite{Garidi-Gazeau-Rouhani-Takook-2008},
Huguet--Queva--Renaud   \cite{Huguet-Queva-Renaud-2008(2)},
Dehghani et al  \cite{Dehghani-Rouhani-Takook-Tanhayi-2008},
Moradi--Rouhani--Takook   \cite{Moradi-Rouhani-Takook-2008},
Faci et al \cite{Faci-Huguet-Queva-Renaud-2009}.

In a number of papers  exact solutions of the wave equations for fields with different spins have been examined
on the background of the de Sitter space and  the problem of
passage of quantum-mechanical particles   (with
different spins  and masses) through the  de~Sitter
horizon:
Lohiya--Panchapakesan \cite{Lohiya-Panchapakesan-1978},
\cite{Lohiya-Panchapakesan-1979},
 Khanal--Panchapakesan \cite{Khanal-Panchapakesan-1981(1)},
 \cite{Khanal-Panchapakesan-1981(2)},
  Khanal \cite{Khanal-1983}, \cite{Khanal-1985},
         Otchik \cite{Otchik-1985}, Motolla
\cite{Motolla-1985}, Bogush--Otchik--Red'kov \cite{Bogush-Otchik-Red'kov-1986},
Mishima--Nakayama \cite{Mishima-Nakayama-1987},
Polarski \cite{Polarski-1989},
Suzuki--Takasugi \cite{Suzuki-Takasugi-1996},
 Suzuki--Taka\-sugi--Umetsu \cite{Suzuki-Takasugi-Umetsu-1998},
 \cite{Suzuki-Takasugi-Umetsu-1999}, \cite{Suzuki-Takasugi-Umetsu-2000}

In particular, a special definition for the {\em reflection} and
{\em transmission} coefficients has been used and  these coef\-ficients have been used to describe
the Hawking radiation on the basis of
the Hawking -- Gibbons formula  \cite{Hawking-1972}, \cite{Hawking-1974}, \cite{Hawking-1975}, \cite{Hawking-1976}
\begin{eqnarray}
{\Gamma _{\epsilon j} \over \exp ( {2\pi E\rho \over \hbar c })  - 1} \; ,
\;\;  \Gamma _{\epsilon j} \; + \; R_{\epsilon j}  = 1 \; ,
\label{Hawking}
\end{eqnarray}
\noindent where $R_{\epsilon j}$  and   $\Gamma _{\epsilon j}$
are  reflection and transmission coef\-fi\-cients, respectively.
Furthermore,  some rather complicated analytical
expressions for these coef\-ficients depending on particle's  spin,
mass, and quantum numbers $(\epsilon, j)$ have been found.

 In
the  paper \cite{Otchik-1985} it was predicted (the case $S=1/2, m
\neq 0$  was considered)  that $R_{\epsilon j} = 0, \; \Gamma _{\epsilon j} = 1 $.
It should be
noted that the main result of the paper  \cite{Otchik-1985},  $R_{\epsilon j} = 0$,
  seems to be an accidental  one in the  sense that the value
$R_{\epsilon j}$ is to be
calculated and in fact it turns to be precisely  zero.

To date, any detailed analysis of the possible explanation for
this  discre\-pancy  has not been given yet.
The present
paper aims  to investigate just these aspects of the
problem.
In the authors'  opinion, the task is  to explore  the key points
in the conventional algorithm for finding the $R_{\epsilon
j}$ and $\Gamma_{\epsilon j}$ in  the~de~Sitter
space-time.

Such
revealing study seems  to be important
  because the general algorithm for
calculation od coefficients  $R$ and $\Gamma$  was taken rather formally
 from more complicated  situations, in particular from the analysis of Schwarzschild  black hole,
where no exact solutions are known:
Starobinskiy --  Churilov \cite {Staroninski-1973}, \cite{Staroninski-Churilov-1973},
Teukolsky -- Press \cite{Teukolsky-1973}, \cite{Press-Teukolsky-1973}, \cite{Teukolsky-Press-1974},
Bardeen -- Press \cite{Bardeen-Press-1973},
Bardeen -- Carter -- Hawking \cite{Bardeen-Carter-Hawking-1973},
Unruh \cite{Unruh-1974},
Fabbri \cite{Fabbri-1975},
Wald \cite{Wald-1975},
Boulware \cite{Boulware-1975},
Page \cite{Page-1976(1)}, \cite{Page-1976(2)}, \cite{Page-1976(3)},
Chandrasekhar -- Detweiler \cite {Chandrasekhar-Detweiler-1977},
Matzner -- Michael \cite{Matzner-Michael-1997},
Guven \cite {Guven-1977},
Bekenstein -- Meisels \cite {Bekenstein-Meisels-1977},
Martellini -- Treves \cite{Martellini-Treves-1977},
Jyer -- Kumar \cite{Jyer-Kumar-1978},
Hawking -- Page \cite{Hawking-Page-1983},
Chandrasekhar \cite{Chandrasekhar}.

Let us outline the content of the paper. In Sec. {\bf 2}, we  briefly
reexamine known  solutions of a scalar field in
the~de~Sitter static coordinates' background. In so doing, certain
 properties relevant for  the subsequent are specified.
In particular, four different solutions are treated in detail (see Sec. {\bf 3})
\begin{eqnarray}
\Psi ^{reg}_{stand}(x)  = e^{-i\epsilon t}
\; f(r) \; Y _{jm} (\theta ,\phi )\;  ,
\nonumber
\\
 \Psi ^{sing}_{stand}
(x) = e^{-i\epsilon t}\; g(r) \; Y_{jm}(\theta ,\phi ) \; ,
\nonumber
\\
\Psi ^{out}_{run}(x) = e^{-i\epsilon t}\;
U^{out}_{run}(r)\; Y_{jm}(\theta ,\phi ) \; ,
\nonumber
\\
\Psi ^{in}_{run}(x) = e^{-i\epsilon t}\; U^{in}_{run}(r)\;
Y_{jm}(\theta ,\phi )\;. \nonumber
\end{eqnarray}

Here,  $f$  and  $g$  are real-valued and linearly independent
solutions of the matter radial equation (see (2.3)),
regular and singular ones at $ r = 0 $ respectively. The functions
$U^{out}_{run}$  and  $U^{in}_{run}$  again are
linearly independent ones, complex-valued, and conjugated
to each other. As a matter of fact, these pairs
$
f(r), \; g(r)  \;\;  \mbox{and}  \;\;
 U^{out}_{run}(r), \; U^{in}_{run}(r)$  can be related to each other by means of  linear
transformations. So, there are some grounds for thinking that
the above solutions $\Phi ^{reg}_{stand}(x)$ and
$\Phi ^{sing}_{stand} (x)$ represent standing waves, whereas
$\Phi ^{out}_{run}(x)$ and $\Phi ^{in}_{run}(x)$ describe
propagating waves.
Besides, as noted in Sec. {\bf 4}, asymptotic behavior
of all these solutions agrees with the used terminology.
In addition, we will demonstrate that as the de~Sitter curvature
radius tends to infinity ($\rho \rightarrow \infty$),
these  solutions are reduced to
the~well known {\em standing} and {\em propagating} waves in the flat
space-time.

In addition, as shown in Sec. {\bf 5}, a radial component
of the scalar particle's conserved current  $J_{r}(x) $
vanishes for the solutions $\Psi ^{reg}_{stand}(x)$ and
$\Psi ^{sing}_{stand} (x)$, whereas
$\Psi ^{out}_{run}(x)$ and $\Psi ^{in}_{run}(x)$
correspond to non-zero currents; moreover, the~relation
$(J_{r})^{in} = - (J_{r})^{out} $ holds.

Finally, it should be added that an explicit form  of the effective
potential  with no barrier  for the~scalar particle's radial
equation (this matter is treated in more detail in Sec. {\bf 6}) points out that there exists no need to
study the process of reflection  in the de Sitter space.
  In the same time, the mere claim that some old  results cannot
be correct seems to be hardly enough.

In sec. {\bf 7} we discuss how approximation in the form of wave functions can influence physical results.
In sec. {\bf 8--10} we extend the results to  the case of electromagnetic field in de Sitter space.
Finally, in sec. {\bf 11--12} we briefly consider the problem for spin 1/2 field in de Sitter
 space\footnote{  The paper is partly a pedagogical review, so we intend to
describe in detail the mathematics  used  to treat particles in the
presence of curved space-time  background.}.

The main goal of the present paper is to clarify the
situation with the Hawking -- Gibbons formula  for the de Sitter space-time with accent on the exact
expression for the coefficient
$\Gamma_{\epsilon j}$ -- it turns to be always (irrespective of spins, masses, and quantum numbers) to be equal
to 1.

To avoid misunderstanding it should be added
explanation for terminology used in the  paper (and in the literature as well).
As a matter of fact, to treat the problem under consideration, we explore the properties of wave functions
in the region far distant from the horizon $r \sim 1$. In so doing,  the quantity
 $R_{\epsilon j} = 1 - \Gamma_{\epsilon j}$  is introduced and called a reflection coefficient,
 though no potential barrier exists in fact.

 However, in the de Sitter space such a potential barrier arises immediately if
 one places at the origin  $r=0$ any point-like  electric charge;  by this reason a
 hydrogen-like atom    is not a stable system in  de Sitter model (see in \cite{preprint-1986})

Though  WKB-calculations are often applied
in treating  Hawking radiation,  quite different  embodiment of the WKB-like idea
was elaborated in  the recent  paper  by Akhmedova --  Pilling --   de Gill --  Singleton \cite{Singleton-2008}.

\section{   Solutions of a radial equation}

The  wave equation\footnote{In essence, we might use an equation without the term 2 at  $M^{2}$,  but this additional
 term makes the wave  equation conformally invariant in the massless case}
for a massive scalar field
$\Phi (x)$    in de~Sitter  space reads
($M = mc\rho /\hbar , \; \rho$ is the curvature radius)
\begin{eqnarray}
( \; {1 \over \sqrt{-g} }  \partial _{\alpha}  \sqrt{-g}
 g^{\alpha \beta }\; \partial _{\beta} +  2  +  M^{2}\;
 ) \; \Psi (x) = 0 \; .
\label{2.1}
 \end{eqnarray}

\noindent  We shall analyze  solutions of this equation in the known static
coordinates
\begin{eqnarray}
dS^{2} =   \Phi  dt^{2} -  {dr^{2} \over \Phi } -
r^{2} (d\theta ^{2} +  \sin ^{2}\theta  d\phi ^{2})   \;,
\nonumber
\\
 0 \le  r \le  1 \; ,  \qquad   \Phi = 1 -r^{2} \;  .
\label{2.2}
\end{eqnarray}

\noindent  Taking $\Psi (x)$  in the~form  of a~spherical wave
\begin{eqnarray}
\Psi (x) = e^{-i\epsilon t} \; f(r) \; Y_{jm}(\theta ,\phi )\; , \qquad \epsilon  = E\rho /\hbar c \; ,
\nonumber
\end{eqnarray}

\noindent
for $f(r)$ we obtain
\begin{eqnarray}
 {d^{2} f \over dr^{2}}   +  ( { 2 \over r} + {\Phi'  \over \Phi } )  {d f\over dr }
    +  \left   (  {\epsilon ^{2} \over \Phi ^{2} }  -
{M^{2} +2 \over   \Phi }  -    {j(j+1) \over   \Phi r^{2}  } \; \right  )
 f   = 0     \; .
\label{2.3}
\end{eqnarray}

\noindent
All solutions of eq.  (\ref{2.3})  can  be  expressed
in terms of hypergeometric functions. To this end, let us introduce
a new variable  $r^{2} = z$, then the equation reads
\begin{eqnarray}
\left [ 4 (1 - z) z {d^{2} \over dz^{2}}  +  (6 - 10 z) {d
\over dz } + {\epsilon ^{2} \over  1 - z} - M^{2} - 2
-  {j(j + 1) \over z} \right ]  f = 0   \; .
\label{2.4}
\end{eqnarray}

\noindent Now, using  the substitution
$
f(z)  = z^{\kappa } (1 - z)^{\sigma } F(z)
$
with parameters $( \kappa  , \; \sigma  )$
given by
\begin{eqnarray}
\kappa  =  j / 2 \; ,\;  - (j + 1)/ 2 \; , \qquad
\sigma  = \pm i \epsilon  / 2 \; ,
\nonumber
\end{eqnarray}

\noindent
for $S(z)$  we obtain
\begin{eqnarray}
z (1 - z) F''\; + \; [ c - (a + b + 1) z ]\; F'\; - \; a b\; F = 0\; ,
\nonumber
\end{eqnarray}

\noindent where $(a , \; b ,\; c)$ satisfy the relations
\begin{eqnarray}
c = 2 \kappa  + 3/2 \;  , \qquad   a + b  =  2 \kappa  + 2 \sigma  + 3/2 \;  ,
\nonumber
\\
a b =  \kappa ^{2} + 2 \kappa  \sigma  + \sigma ^{2} +
{3 \over 2} \kappa
+ {3 \over 2} \sigma  + {M^{2} + 4 \over 2}  \; .
\nonumber
\end{eqnarray}

\noindent Taking $ \kappa  = j/2$ and $\sigma =  -i\epsilon /2$,
 we get a regular (at  $r  =  0 $)   solution
\begin{eqnarray}
\kappa = j/2 \; ,\; \;  \sigma = -i\epsilon /2\;,  \;\;  c = j + 3/2 \;,
\nonumber
\\
f(z) = z^{j/2} \; (1 - z)^{-i\epsilon /2}\; F( a , b , c ; z ) \;,
\nonumber
\\
a =  { 3/2 + j + i \sqrt{ M^{2}-1/4}  - i\epsilon  \over 2} \; ,
\nonumber
\\
  b =  { 3/2 + j - i \sqrt{ M^{2}-1/4}  - i\epsilon  \over 2} \; .
\label{2.5b}
\end{eqnarray}

\noindent
 In turn, choosing  $ \kappa = -(j + 1)/ 2$ and $\sigma =  - i\epsilon / 2 )$,  we  obtain a
singular (at  $r  =  0 $) solution
\begin{eqnarray}
\kappa  = -(j + 1)/2 \; , \;  \sigma  = -i\epsilon /2 \; , \; c = -j + 1/2 \; ,
\nonumber
\\
g(z)  = z^{-(j+1)/2}\; (1 - z)^{-i\epsilon /2} \; F( \alpha
,\;\beta ,\; \gamma ; \; z ) \; ,
\nonumber
\\
\alpha  =  { 1/2 - j + i \sqrt{ M^{2}-1/4}  - i\epsilon  \over 2}
\; ,
\nonumber
\\
 \beta = { 1/2 - j - i \sqrt{ M^{2}-1/4}  - i\epsilon
\over 2} \; .
\nonumber
\\
 \label{2.6b}
 \end{eqnarray}

\section{Standing and propagating waves}

It is easily verified that the regular solution above
represents a real-valued function. Besides,  it can be expanded into superposition of two
complex-valued (and conjugate) solutions of
the same equation (\ref{2.3}).     Applying  the terminology used in
the ordinary case of spherical waves in the flat space-time model,
one can say that the regular standing   wave is a certain
superposition of two propagating waves. One propagating wave goes to the
de~Sitter horizon (outgoing wave), another runs backwards (ingoing wave).
The same is valid for the singular standing wave: the $g(r)$ is
a~real-valued one, and it can be expressed as a~linear
combination of the~same complex-valued solutions (with
other coefficients).

Now, let us examine this matter in  detail. First let us turn to  the regular solution.
 To get a required decomposition,
it is sufficient to use one of the so-called Kummer's
relationships
\begin{eqnarray}
U_{1} =  { \Gamma (c) \Gamma (c - a - b) \over \Gamma (c - a) \Gamma (c - b)} \;
U_{2} \; + \;
{ \Gamma (c) \Gamma (a  + b  -c ) \over \Gamma (a) \Gamma (b) }\; U_{6}
  \; ,
\label{3.1a}
\end{eqnarray}

\noindent where $U_{1} ,\; U_{2} ,\; U_{6}$  represent three
different solutions of the same hypergeometric equation determined by
\begin{eqnarray}
U_{1}  = F (a , b , c ; z ) \; , \;
U_{2}  = F (a , b , a + b - c + 1 ; 1 - z )\; , \;
\nonumber
\\
U_{6}  = (1-z)^{c-a-b} F ( c - a , c - b , c - a - b + 1 ; 1 - z )
\; . \label{3.1b}
\end{eqnarray}

\noindent Applying the relation  (\ref{3.1a})  to the solution $f(z)$
as $U_{1}$, we  get
\begin{eqnarray}
f(z) =  { \Gamma (c) \Gamma (c - a - b)  \over
\Gamma (c - a) \Gamma (c - b) } \; U^{out}_{run}(z) \; + \;
{\Gamma (c) \Gamma (a + b - c ) \over \Gamma (a) \Gamma (b) }\;
 U^{in}_{run}(z) \; ,
\label{3.2a}
\end{eqnarray}

\noindent where
\begin{eqnarray}
U^{out}_{run}(z)  = z^{j/2} \; (1 - z)^{-i\epsilon /2} \;
F ( a , b , a + b - c  + 1 ; 1  -  z ) \;  , \;\;
\nonumber
\\
U^{in}_{run}(z)  = z^{j/2}\; (1- z)^{+i\epsilon /2}\;
 F (c- a, c - b, c - a - b + 1; 1 - z) \; .
\label{3.2b}
\end{eqnarray}

\noindent Noting that
\begin{eqnarray}
a^{*} = ( c - a ) \; ,\;  b^{*} = ( c - b ) \;  , \; ( a + b - c
)^{*}  =  - ( a + b - c ) \; , \nonumber
\end{eqnarray}
 one can conclude that  two functions
$U^{out}_{run}(z)$  and   $U^{in}_{run}(z)$  are
related to each other by  complex conjugation
\begin{eqnarray}
[\; U^{out}_{run}(z) \; ]^{*}   =  U^{in}_{run}(z)\;  .
\nonumber
\label{3.3b}
\end{eqnarray}

\noindent Besides, the function  $f(z)$ being a sum of two
complex-conjugate expressions  (see (\ref{3.2a}) and (\ref{3.3b})) is
a~real-valued one. Hence, one gets
\begin{eqnarray}
f(z)  = 2\; \mbox{Re}\;  {\Gamma (c) \Gamma (c - a - b) \over
\Gamma (c - a) \Gamma (c - b)} \; U_{out}(z)\;
\nonumber
\\
= 2\; \mbox{ Re}\;
\;{ \Gamma (c) \Gamma (a  + b  - c ) \over \Gamma (a)
\Gamma (b)} \; U_{in}(z) \; \;  .
\label{3.3c}
\end{eqnarray}

To obtain an analogous expansion for the singular function
$g(z)$, one has to apply the same Kummer's  formula
(\ref{3.1a}), but now replacing  $(a,\; b,\; c)$ by
 $(\alpha ,\;\beta , \; \gamma )$
\begin{eqnarray}
U_{1} =   { \Gamma (\gamma ) \Gamma (\gamma - \alpha  -
\beta ) \over \Gamma (\gamma - \alpha ) \Gamma (\gamma  - \beta )
}\; U_{2} \; +  \; { \Gamma (\gamma ) \Gamma (\alpha + \beta -
\gamma  ) \over \Gamma (\alpha ) \Gamma (\beta ) } \; U_{6}
 \; ,
\label{3.4a}
\end{eqnarray}

\noindent where
\begin{eqnarray}
U_{1} = F ( \alpha  , \beta  , \gamma  ; z ) \; , \;
U_{2} = F ( \alpha  , \beta  , \alpha  + \beta  - \gamma  + 1 ; 1 - z ) \; ,
\nonumber
\\
U_{6} = (1-z)^{\gamma - \alpha - \beta} F ( \gamma  - \alpha  ,
\gamma - \beta , \gamma - \alpha  - \beta  + 1 ; 1 - z )  \;  .
\label{3.4b}
\end{eqnarray}

\noindent Thus we arrive at
\begin{eqnarray}
g(z) =   {\Gamma (\gamma ) \Gamma (\gamma  - \alpha  -
\beta ) \over \Gamma (\gamma  - \alpha ) \Gamma (\gamma  - \beta
)}\; U^{out}_{run}(z) \; + \; {\Gamma (\gamma ) \Gamma (\alpha  +
\beta  - \gamma ) \over \Gamma (\alpha ) \Gamma (\beta ) }\;
U^{in}_{run}(z) \; ,
\label{3.5a}
\end{eqnarray}

\noindent where
\begin{eqnarray}
U^{out}_{run}(z) = z^{j/2} \; (1- z)^{-i\epsilon /2} F ( \alpha
+ 1 - \gamma , \beta  + 1 - \gamma , \alpha  + \beta  + 1 - \gamma
; 1  - z )\;  ,
\nonumber
\end{eqnarray}
\begin{eqnarray}
U^{in}_{run}(z) = z^{j/2}\; (1- z)^{+i\epsilon /2} F ( 1 -
\alpha, 1 - \beta ,\gamma + 1 - \alpha - \beta ;1 - z)\; .
\label{14.3.5c}
\end{eqnarray}

\noindent The  following relations
\begin{eqnarray}
\alpha  + 1 - \gamma   = a \; , \;
\;  \beta  + 1 - \gamma   = b \; ,
\nonumber
\\
 \alpha  + \beta  - \gamma  + 1  =  a + b + 1 - c \; ,
\nonumber
\\
1 - \alpha   = c - a  \;  , \; 1 - \beta   = c - b
\nonumber
\end{eqnarray}

\noindent are taken into account. Again, the function
$g(z)$ is real-valued one
\begin{eqnarray}
g(z)  = 2 \; \mbox{Re}\; {\Gamma (\gamma ) \Gamma (\gamma -
 \alpha - \beta) \over
 \Gamma (\gamma - \alpha ) \Gamma (\gamma - \beta )}  \;
 U^{out}_{run}(z)
 \;
 \nonumber
 \\
  = 2 \; \mbox{Re} \;  {\Gamma (\gamma ) \Gamma (\alpha + \beta  - \gamma  ) \over \Gamma
(\alpha ) \Gamma (\beta ) } \; U^{in}_{run}(z)\;    .
\label{3.5d}
\end{eqnarray}

\section{ Asymptotic behavior and flat space-time limit}

Now let us consider how the above four solutions behave at  the
origin  $r \sim 0$ and at the horizon  (when  $r \sim \; 1$).
First, for  $U^{out}_{run}$ we have
\begin{eqnarray}
U^{out}_{run}(z) = z^{j/2} \; (1 - z)^{-i\epsilon /2} \; F ( a ,
b , a + b - c + 1; 1 - z ) \; .
\label{4.1a}
\end{eqnarray}

\noindent Again, applying the formula (\ref{3.1a})  to $U^{out}_{run}(z)$
as  $U_{1}$, we arrive at the following expansion
\begin{eqnarray}
U^{out}_{run}(z) =  {\Gamma (a + b + 1 - c)\Gamma (1 -
c) \over \Gamma (b - c + 1) \Gamma (a  - c + 1) } \; z^{j/2}\; (1
- z)^{-i\epsilon /2} \;
F ( a , b , c ; z ) \;
\nonumber
\\
+\; { \Gamma (a + b + 1 - c) \Gamma (c -1) \over \Gamma (a) \Gamma (b) }
\; z^{-(j+1)/2}\; (1 -z)^{-i\epsilon /2} \;
\nonumber
\\
\times
 F (b - c + 1,
a - c + 1,- c + 2; z).
\label{4.1b}
\end{eqnarray}

\noindent The last is appropriate for asymptotic analysis, so one gets
\begin{eqnarray}
U^{out.}_{run} ( r \sim  0 ) \sim  { 1 \over r^{j+1}} \; , \qquad
U^{out}_{run}( r \sim  1 ) \sim  (1 - r^{2})^{-i\epsilon /2}\; .
\label{4.2a}
\end{eqnarray}

\noindent Introducing a new radial variable $r^{*}$
\begin{eqnarray}
r^{*}  = {\rho  \over 2} \; \ln  {1 + r \over 1-r } \;\; , \qquad
r = {\exp (2r^{*}/\rho ) - 1 \over  \exp (2r^{*}/\rho ) + 1 } \;\;
, \qquad  r \in  [ 0,\; \infty  ) \; , \nonumber
\end{eqnarray}

\noindent one can easily express the second asymptotic
relation in (\ref{4.2a}) in the form $(\epsilon = E\rho /\hbar c )$
\begin{eqnarray}
U^{out}_{run}( r^{*} \sim  \infty  ) \sim \left (
 2^{-iE\rho /\hbar c} \right ) \;  \exp (+i E r^{*}/\hbar c )\; .
\label{4.2b}
\end{eqnarray}

\noindent Analogously, one can find asymptotics for the in-wave
\begin{eqnarray}
U^{in}_{run}( r \sim 0 ) \sim  {1 \over r^{j+1} }\; ,  \qquad
U^{in}_{run}( r\sim 1) \sim (1 - r^{2})^{+i\epsilon /2}
\label{4.3a}
\end{eqnarray}

\noindent or
\begin{eqnarray}
U^{in}_{run}( r^{*} \sim  \infty  ) \sim  \left ( 2^{+iE\rho
/\hbar c} \right ) \; \exp ( -iE r^{*} /\hbar c ) \; .
\label{4.3b}
\end{eqnarray}

\noindent Proceeding with the analysis for the standing waves,
 we   get
\begin{eqnarray}
f(r \sim  0) \sim  r^{j}\; ,  \qquad
g(r \sim  0) \sim { 1 \over
r^{j+1}} \; ,
\label{4.4a}
\end{eqnarray}
\begin{eqnarray}
f(r \sim 1 ) \sim    \left [\;
{\Gamma (c)\Gamma (c-a-b) \over  \Gamma (c - a) \Gamma (c - b)} \;
( 1 - r^{2} )^{-i\epsilon /2}  \; \right.
\nonumber
\\
\left.
+\;
{ \Gamma (c) \Gamma (a + b - c)  \over \Gamma (a) \Gamma (b) }
\;  ( 1 - r^{2} )^{+i\epsilon /2} \; \right ]
\nonumber
\\
= 2\; \mbox{Re} \; \left [ \; { \Gamma (c) \Gamma (a  + b - c ) \over
\Gamma (a) \Gamma (b) }\; 2^{+iE \rho /\hbar c} \; \mbox{exp}(-iE r^{*}
/\hbar c) \;    \right ] ,
\label{4.4b}
\end{eqnarray}
\begin{eqnarray}
g(r \sim 1) \sim   \left  [  \;
{\Gamma (\gamma ) \Gamma (\gamma - \alpha - \beta ) \over
\Gamma (\gamma  - \alpha ) \Gamma (\gamma  - \beta )} \;
(1- r^{2})^{-i\epsilon /2} \;
\right.
\nonumber
\\
\left.
+\; { \Gamma (\gamma )\Gamma (\alpha
+\beta -\gamma ) \over  \Gamma (\alpha ) \Gamma (\beta )}\; ( 1 -
r^{2} )^{+i\epsilon /2}  \; \right ]
\nonumber
\\
= 2 \; \mbox{Re}  \;\left [\; { \Gamma (\gamma ) \Gamma (\alpha  + \beta  -
\gamma ) \over \Gamma (\alpha ) \Gamma (\beta )  } \; 2^{+iE \rho
/\hbar c} \; \mbox{exp}(-iE r^{*} /\hbar c) \; \right ] .
\label{4.4c}
\end{eqnarray}

Now, let us consider  a limiting transition to the flat space-time. For definiteness we take
the   outgoing wave. It will be convenient to start with its
representation according to (\ref{4.1b}).

Then
rewriting  the above expressions for the parameters  $(a ,\; b ,\;
c)$ with a curvature radius $\rho$  (a subsequent limiting procedure will be realized just through
the limit $\rho\; \rightarrow  \; \infty  $ )
\begin{eqnarray}
a = {p + 1 - i \epsilon  \rho  + i \sqrt{\rho ^{2} M^{2} - 1/4} \over
2} \; ,
\nonumber
\\
b = {p + 1 - i \epsilon \rho -i \sqrt{\rho ^{2}M^{2} -1/4}\over 2}\; ,
\nonumber
\\
(a - c + 1) = { - p + 1 - i \epsilon  \rho  + i\sqrt{ \rho ^{2}
M^{2} - 1/4} \over 2} \; ,
\nonumber
\\
(b - c + 1) = { - p + 1 - i \epsilon
\rho  - i \sqrt{ \rho ^{2} M^{2}  - 1/4} \over 2}\; ,
\nonumber
\\
p = j + 1/2 \;\; ; \;\; ( a + b + c ) = ( 1 - i \epsilon  \rho  )
\; ,
\label{4.5}
\end{eqnarray}

\noindent and
$\lim_{ \rho \rightarrow  \infty } ( \rho ^{2} z ) = R^{2}\;$, where
$R$ is the usual radial coordinate of the flat space-time.

Using a  series representation for a~hypergeometric
function (when $ \mid z\mid  < 1 $)
\begin{eqnarray}
F ( a , b , c ; z)=\left [ 1  + { a b \over c } {z \over 1!}\;+\;
{ a(a+1) b(b+1) \over c(c+1) } \; {z^{2} \over 2!}\; + \; \ldots
\right ] \; ,
\label{4.6a}
\end{eqnarray}

\noindent and further taking into account two relations
\begin{eqnarray}
\lim_{ \rho \rightarrow \infty } \left [\; {(a +n)(b +n) \over \rho
^{2} }\; z^{2} \rho ^{2} \; \right ]  =
- (\epsilon ^{2} - M^{2}) \; { r^{2} \over 4 } \; ,
\nonumber
\\
\lim _{\rho \rightarrow \infty } \left [  \;
{(a-c+1)(b-c+1) \over \rho ^{2}} \; z^{2} \rho ^{2} \;\right ] = -
(\epsilon ^{2} - M^{2}) \;{ r^{2} \over 4 } \; ,
\nonumber
\end{eqnarray}

\noindent one comes to ($ \epsilon ^{2} - M^{2} \equiv k^{2} $)
\begin{eqnarray}
\lim _{ \rho \rightarrow \infty } F (a , b , c ; z ) = \Gamma (1 +
p)\; \sum_{0}^{\infty} {(- k^{2} r^{2}/4)^{n} \over n! \Gamma (1
+ n  + p) }\; ,
\nonumber
\label{4.6b}
\\
\lim_{ \rho \rightarrow \infty} F(b - c + 1, a - c + 1, - c + 2 ;
z)
\nonumber
\\
 = \Gamma (1 - p) \; \sum_{0}^{\infty} {(-k^{2} r^{2}/4)^{n}
\over n!  \Gamma (1 + n - p)}\; .
\label{4.6c}
\end{eqnarray}

\noindent Now, with the use of known expansion for the~Bessel
function
\begin{eqnarray}
J_{p}(x) = ({x \over 2})^{p} \; \sum _{0}^{\infty} {(ix/2)^{2n} \over
n! \Gamma ( 1 + n + p)}
\nonumber
\end{eqnarray}

\noindent from  (\ref{4.6b})--(\ref{4.6c}) one gets
\begin{eqnarray}
\lim _{ \rho \rightarrow \infty  } \left [\; \rho ^{j}\; r^{j}\;
(1 - r^{2})^{-i\epsilon /2} \; F (a, b, c; z)\; \right ]  = ({2
\over k})^{+p} \; \Gamma (1 + p) \; {J_{p}(kr) \over \sqrt{r}}\; ,
\nonumber
\\
\lim_{ \rho \rightarrow \infty } \left [\; \rho^{-j-1} \; r^{-j-1}
\; (1-r^{2})^{-i\epsilon /2} \; ({2 \over k})^{-p} \; \Gamma (1 -
p)\; {J_{-p}(kr) \over \sqrt{r}} \; \right ] \; .
\nonumber
\\
\label{4.7b}
 \end{eqnarray}

\noindent Further, taking into account  (\ref{4.5})  and   (\ref{4.7b}), for
$U^{out}_{run}$  at the limit  $\rho \rightarrow  \infty $
we obtain   ($A$ is a  constant)

\begin{eqnarray}
\lim _{ \rho  \rightarrow \infty } A\; U^{out}_{run}(z) =
\lim_{ \rho \rightarrow  \infty } A \;  {1 \over \sqrt{r}}
\nonumber
\\
 \left [ {\Gamma  (-i\epsilon \rho + 1) \Gamma(-p) \;
\Gamma (1+p)  (2/k)^{p}  J_{p}(kr)  \rho^{-p+1/2} \over
\Gamma [{1 \over 2} (+i \sqrt{\rho ^{2} M^{2} -1/4 } - i\epsilon \rho
+ p + 1) ]  \Gamma [{1\over 2} (-i \sqrt{\rho^{2}
M^{2}  - 1/4} - i \epsilon \rho + p +1)]} \right.
\nonumber
\\
\left.   + {\Gamma (-i\epsilon \rho +1) \Gamma(+p)  \Gamma (1-p)
 (2/k)^{-p}  J_{-p}(kr)  \rho ^{+p+1/2} \over \Gamma [{1
\over 2} (+i \sqrt{\rho^{2} M^{2} -1/4} - i\epsilon \rho -p +1)]
\Gamma [{1 \over 2} (-i \sqrt{\rho^{2} M^{2} -1/4} - i\epsilon
\rho -p + 1)]}  \right]    .
\nonumber
\label{4.8}
\end{eqnarray}

\noindent Now, using the known relation for $\Gamma$-function
\begin{eqnarray}
\Gamma (p) \; \Gamma (1 - p) = { \pi \over \sin (\pi
p)} \; ,
\nonumber
\end{eqnarray}

\noindent and introducing  the notation
\begin{eqnarray}
{- i \epsilon \rho  + i \sqrt{\rho ^{2} M^{2} - 1/4 } \over 2}
\sim   i\;\rho\;  {M - \epsilon \over 2}  = \rho  z_{1} \;  ,
\nonumber
\\
{- i \epsilon \rho   - i \sqrt{\rho ^{2} M^{2}  - 1/4 } \over 2}
\sim  i\;\rho \;  {M + \epsilon \over 2} = \rho z_{2} \; ,
\nonumber
\end{eqnarray}

\noindent we transform  the above relation  into
\begin{eqnarray}
\lim_{\rho \rightarrow \infty} A\; U^{out}_{run}(z) =
 \lim_{\rho \rightarrow \infty } A\; {\pi \over \sin (\pi p)} \;
{ \Gamma (- i \epsilon  \rho  + 1) \over \sqrt{r}}
\nonumber
\\
\times
\left [ \;
- {(2/k)^{p} \; J_{p}(kr)\; \rho ^{-p+1/2} \over
\Gamma(\rho z_{1}  - {1-p \over 2}) \; \Gamma (\rho z_{2}-{1-p\over
2}) } \; + \;( p \; \rightarrow \;  - p)\; \right ] \;
\label{4.9}
\end{eqnarray}

\noindent  the designation $( p \; \rightarrow \;  - p)$ means that
 a  term is to be placed there (its form follows  from  the previous one through the change $p \; \rightarrow \;  - p$).
The constant $A$ will be chosen by convenience reason later; so it is
\begin{eqnarray}
\lim _{ p \rightarrow  \infty  } A  = \lim_{ p \rightarrow \infty
} {\Gamma (a - {c-1 \over 2}) \; \Gamma (b - {c-1  \over 2}) \over
\Gamma ( a + b - c + 1 )}  = \lim _{ p \rightarrow  \infty  }
{\Gamma (\rho z_{1} + 1/4) \; \Gamma (\rho z_{2} + 1/4) \over
\Gamma (-i\epsilon \rho +1) }\; . \label{4.10}
\end{eqnarray}

\noindent Performing  calculation for  the first
term in (\ref{4.9}) (it will be noted  as  $F(p)$)
the contribution of the second term  is determined through the  change $p \rightarrow - p$)
\begin{eqnarray}
F(p) = \lim_{p \rightarrow \infty } \left  [ {-\pi  \over \sin
(\pi p)}   {1 \over \sqrt{r} } \; {(2/k)^{p}  J_{p}(kr) \over 2^{p
-1/2} }  {\Gamma (\rho z_{1}+ 1/4) \over \Gamma (\rho z_{1} - {1-p
\over 2})}  {\Gamma (\rho z_{2} + 1/4) \over \Gamma (\rho z_{2} -
{1-p\over 2})}  \right ]  . \label{4.11a}
\end{eqnarray}

\noindent  Applying   the~above formula referring  $\Gamma$-s with arguments  $p$
and  $(1 - p)$ again, one  produces
\begin{eqnarray}
F(p) = \lim _{p \rightarrow \infty } \; {-\pi \over \sin
(\pi p)}\; {1 \over \sqrt{r} }\; {(2/k)^{p}\; J_{p}(kr) \over 2^{p
-1/2} } \;
\nonumber
\\
\times  {\Gamma (\rho z_{1}+ 1/4) \over \Gamma (\rho z_{1} + {1
- p \over 2})} \; {\sin  \pi (\rho z_{2} + {1-p \over 2}) \over
\sin  \pi (\rho z_{2} + 1/4) }  \; {\Gamma (-\rho z_{2}  + {1+p
\over 2}) \over \Gamma (-\rho z_{2} + 3/4)} \; ,
\label{4.11b}
\end{eqnarray}

\noindent which after the use of the known asymptotic
relationship for $\Gamma$-function gives
\begin{eqnarray}
 x \; \rightarrow \; \infty \; , \; \mid \arg  x\mid  < \pi \;
  \rightarrow \; { \Gamma (x + \alpha ) \over \Gamma (x + \beta ) }
  \sim  x^{\alpha -\beta }.
\nonumber
\end{eqnarray}

\noindent  Taking  into account  the  identity
\begin{eqnarray}
 { \sin  \pi ( \rho z_{2} + {1-p \over 2} ) \over \sin  \pi (
  \rho z + 1/4 ) }
  \nonumber
  \\
  = { \exp\;  [\pi  {\rho (M + \epsilon ) \over 2}
\;+\; i \pi { 1+p \over 2}] - \exp \;  [-\pi  {\rho (M + \epsilon )
\over 2}\; -\; i \pi { 1+p \over 2}] \over \exp \; [ \pi {\rho (M +
\epsilon ) \over 2} \;+\; i \pi /4  ] - \exp \; [-\pi {\rho (M +
\epsilon ) \over 2} \;-\; i \pi / 4 ] }
 \nonumber
 \\
 = \lim_{ \rho \rightarrow
 \infty } \exp \; [i \pi ({p \over 2} + {1 \over 2} ]\; ,
\nonumber
\end{eqnarray}

\noindent we obtain  the formula
\begin{eqnarray}
 F(p) = \lim _{ \rho \rightarrow \infty } t
[ \; {-\pi \over \sin \pi p }  \; {1 \over \sqrt{r}} \; ({2 \over
k})^{p}\;
\nonumber
\\
\times
J_{p}(kr) \; \exp\;  [ ( \rho  z_{1})^{p/2-1/4}\; ( \rho
z_{2})^{p/2 -1/4} ] \; \exp \; [ i \pi ({p \over 2} + {1 \over 2})]
\; ,
\nonumber
\end{eqnarray}

\noindent from whence it follows that
\begin{eqnarray}
F(p) =  \sqrt{{2 \over kr}}\; \exp \; [i \pi ({p \over 2} + {1 \over
2})] \; {\pi \over \sin  \pi p } \; J_{p} ( kr) \; .
\label{4.11c}
\end{eqnarray}

\noindent Allowing for the second term's contribution ($p \rightarrow
- p$), the~final expression for  $A \; U_{out}(z)$ has the form
\begin{eqnarray}
\lim _{ \rho \rightarrow  \infty } A \; U_{out}(z)  = {1 \over
i^{j+1} } \;  \sqrt{{2 \over kr}} \; H^{(1)}_{j+1/2}(kr) \; ,
\label{4.12a}
\end{eqnarray}

\noindent where $H^{(1)}_{j+1/2}(kr)$  denotes the~Hankel function
\begin{eqnarray}
 H^{(1)}_{j+1/2}(x) = { ip \over \sin (\pi p)}
\; \left  [ \; e^{ip\pi } \; J_{p}(x) \; - \;
J_{-p}(x)\;\right]\;.
\label{4.12b}
\end{eqnarray}

\noindent Eq. (\ref{4.12a}) provides us with the conventional
representation for an~expanding spherical wave
 in   flat space-time.

In connection with this  limiting procedure   to flat
space-time, an important  point
should  be  stressed. Because   the~curved  metrics
(\ref{2.2})  in fact coincides  with the~flat one as the region is sufficiently far  from the horizon ($ 0 \leq r << \rho $), one
might expect a little difference between {\em curved } and {\em flat} \hspace{2mm}
solutions of the~respective  (scalar particle's) equations:
$
f_{curve}(r)  \approx  f_{flat}(r)  \; .
$
This statement can be formulated mathematically as
\begin{eqnarray}
A \;U^{out}_{run}(z)  \; \;  \longrightarrow \;\;   { 1 \over i^{j + 1} }  \; \sqrt{{2
\over kr}}\; H^{(1)}_{j + 1/2} (kr)\; .
\label{4.13}
\end{eqnarray}

 However, the relation (\ref{4.13}) imposes several  additional constraints  on parameters.
 Indeed,  let us compare  two
equations
\begin{eqnarray}
\mbox{in \;  de \; Sitter \;model },  \qquad
\Phi  = e^{-i\epsilon t} \; f_{\epsilon j}(r) \;
Y_{jm}(\theta ,\phi) \; ,
\nonumber
\\[3mm]
\left [{d^{2}\over dr^{2} } \;+\; {2(1 - 2r^{2}) \over r(1 -
r^{2})}\; {d \over dr}\; +\; {\epsilon ^{2} \over (1 - r^{2})^{2}
}\; -\; {M^{2}  + 2 \over 1 - r^{2} } \; -\; {j(j + 1) \over
r^{2}} \right ]  f_{\epsilon j}  = 0 \; ;
\nonumber
\\
\label{4.14}
\end{eqnarray}
\begin{eqnarray}
\mbox{in\; Minkowski \; model }, \qquad
\Phi ^{0}= e^{-i\epsilon t} \;
 f^{0}_{\epsilon j}(r) \; Y_{jm}(\theta ,\phi ) \; ,
\nonumber
\\[3mm]
\left [ {d^{2}\over dr^{2} } \;+\; {2 \over r } {d \over dr} \;+\;
\epsilon ^{2} \;-\; M^{2} \; - \; {j(j + 1) \over r^{2}} \right ]
 f_{\epsilon j} ^{0} = 0 \; .
\label{4.15}
\end{eqnarray}

\noindent In the region  $r << \rho $, the equation (\ref{4.14}) reads
\begin{eqnarray}
\left [ \; {d^{2}\over dr^{2} } \;+\; {2 \over r } {d \over dr} \;
( 1 - {r^{2} \over \rho ^{2}} + \ldots ) \;+\; \epsilon ^{2}\;
( 1 + {2 r^{2} \over \rho ^{2}} + \ldots ) \;   \right.
\nonumber
\\
\left. -\; ( M^{2} + {2 \over \rho^{2}})(1 + {r^{2} \over \rho^{2}} +
\ldots ) \;-\; {j(j + 1) \over r^{2}} \; ( 1 + { r^{2} \over
\rho^{2}} \; +\; \ldots ) \; \right ]  f_{\epsilon j} = 0 \; .
\label{4.16}
\end{eqnarray}

\noindent Eqs. (\ref{4.16}) and  (\ref{4.15})   coincide only  if   two restrictions hold
\begin{eqnarray}
 r << \rho , \qquad  \mbox{and}  \qquad
 \epsilon ^{2} - M ^{2} - {j(j + 1) \over r^{2}}
  >>  {j(j + 1) \over \rho ^{2}} +
 {2 \over \rho^{2}}  \; ;
\label{4.17}
\end{eqnarray}

\noindent from whence it follows that
\begin{eqnarray}
\epsilon ^{2} \; - \; M ^{2}  >> { j ( j + 1)  +  2 \over
\rho^{2} }\;  .
\label{4.18}
\end{eqnarray}

\noindent  Therefore, the~difference between a~particle  energy
$\epsilon$ and mass $m$ should be  big  enough in comparison with
the  parameter  $(j/\rho )$. Here  $\rho $ denotes the curvature
radius, and  $j$ is an angular momentum number. At  $j$ = 0, the
relation $(\epsilon ^{2}$ - $M^{2} ) >> 2/\rho ^{2}$ ought to imposed.  The curvature radius  $\rho $ is  large
(but finite), whereas   $j=0, 1, 2, ..., n, ...$ is unbounded.

\section{  Standing, propagating waves and explicit expression\\  for  conserved current}

As an additional argument for  the above-used terminology,
let us consider  properties of the  particle's  conserved
current
\begin{eqnarray}
J_{\alpha}(x) = i \; ( \Phi ^{*} \nabla _{\alpha } \Phi \; -  \; \Phi
^{*} \nabla_{\alpha } \Phi ) \;.
\label{5.1}
\end{eqnarray}

\noindent Using  general representation of the spherical wave
$
\Phi (x) = e^{-i\epsilon t} f(r) Y_{jm}(\theta ,\phi ),
$
for current's components $J_{\alpha }(x)$ one gets
\begin{eqnarray}
J_{t} = 2 \epsilon  \mid  f \mid ^{2} \; \mid  \Theta _{jm}\mid
^{2}\; , \;\; J_{\phi }  = - 2 m \; \mid  f \mid ^{2}\;
 \mid \Theta_{jm}\mid ^{2} \; ,
\nonumber
\\
J_{r} = i ( f^{*} {d \over dr}f \;-\; f{d \over dr}f^{*} ) \; \mid
\Theta _{jm}\mid ^{2} \; ,\; \; J_{\theta } = 0\; .
\label{5.2}
\end{eqnarray}

\noindent From whence it follows that standing and propagating solutions
 are characterized  by
\begin{eqnarray}
(J_{r})^{reg}_{stand} \equiv 0 \, ,  \;\;
(J_{r})^{sing}_{stand} \equiv 0 \, ,     \;\;
(J_{r})^{out}_{run} = - \; (J_{r})^{in}_{run} \; .
\nonumber
\end{eqnarray}

\noindent  These properties of the  radial  component  of the current are in
agreement with the notation used:
$\Phi^{out(in)}_{run}$  is a  propagating wave,
$\Phi^{reg(sing)}_{stand}$ represents  a standing wave.

 \section{ On transparency  of
the de Sitter horizon }

From the preceding material it might be concluded that the
phenomenon itself of any {\em reflection of scalar particles by
the~Sitter event horizon} should not be observed whatsoever. Additional
credence to that opinion may be given by  consideration of the problem in terms of
the one-dimensional Schr\"{o}dinger's like equation with an~effective
potential $U(r)$. Indeed, the~matter radial  equation (\ref{2.3}) can be
taken into the form (the  variable $r^{*}$ is used again)
\begin{eqnarray}
\left [ {d^{2} \over dr^{*2}}  +   \epsilon ^{2} - U(r^{*})
\right ]  G(r^{*}) = 0 \; ;
\label{6.1a}
\end{eqnarray}

\noindent where  $G(r^{*})$ is introduced by
\begin{eqnarray}
f(r)  = \mu (r^{*})  \; G(r^{*})\; , \qquad
 {d \over dr^{*}} \; \ln  \mu = { (1 - r)(1 + 2r) \over \rho }
 \; ,
\nonumber
\end{eqnarray}

\noindent and the effective potential $U(r^{*})$ is
determined as
\begin{eqnarray}
U(r^{*}) =    {1 - r^{2} \over \rho ^{2}} \left [ 4 (1 - r) \;+\;
{r \over 1 + r} \; + \; m^{2} \rho^{2} \;+\; {j (j + 1) \over
r^{2}}    \right ] \; .
\label{6.1b}
\end{eqnarray}

\noindent As readily verified, the potential (\ref{6.1b}) corresponds to an
effective repulsive force  $ - d U /  dr^{*}  $ at every
spatial point  $( r \le 1 ; r^{*} \in [ 0, +\infty ])$ of the
de Sitter space-time. In other words, the center $r = 0$  is effectively
repulsive for a scalar particle everywhere and no barrier
appears between  the~center and event
horizon $r \sim \rho$
\begin{eqnarray}
F_{r^{*}} \equiv    - {d U \over dr^{*} } =
{1 - r^{2} \over \rho ^{2}} \;+\; \left [ 2 r \left (
{j (j + 1) \over r^{2} }\;+\; m^{2} \rho ^{2} + \;
{r\over 1 + r}
\; \right.  \right.
\nonumber
\\
\left.  \left.  +\; 4 (1 - r) \right ) \;+\;   (1 - r^{2}) \left (
{2 j (j + 1) \over r^{3} } \;+\; 4 - {1 \over (1 + r)^{2}}
\right )
\right ] > 0 \; .
\nonumber
\end{eqnarray}

\noindent In the region near to  horizon ($r^{*} \sim + \infty)$,
the~potential  $U(r^{*})$ tends to zero, and
correspondingly the solutions (propagating waves are meant here for
definiteness) look like exponents
\begin{eqnarray}
G(r^{*}) \sim  \exp  ( \pm  i \epsilon r^{*}) \; ;
\nonumber
\end{eqnarray}

\noindent the  boundary behavior may be naturally
interpreted as associated with two waves passing to ($+$) and from  ($-$)
the~de~Sitter horizon respectively.

After  examination of this Scr\"{o}dinger's-like equation, it is clear  that the quantity such as {\em a
reflection coefficient $R_{\epsilon j}$} cannot be correctly
determined.  However,  just to make this statement is  not enough, because   a bit of vagueness would still
remains.

To clarify the subject,
now let us reexamine  an algorithm  applied in the literature for
calculating of such a~quantity  as $R_{\epsilon j}$ on the~background of
the~de~Sitter space-time model. For definiteness and simplicity,  let us consider the~case of
massless particles.

A wave passing to the~horizon  is described by the following radial
function (see (\ref{4.1b})  at  $m^{2} = 0$  and  $r = R/\rho$)
\begin{eqnarray}
U^{out}_{run}(R) =
{ \Gamma (a + b + 1 - c) \;\Gamma (1 - c) \over
\Gamma (b - c + 1)\; \Gamma (a - c + 1)} \;
r ^{j}\; (1 - r^{2})^{-i\epsilon \rho /2} \;
F ( a, b , c ; r^{2})  \;
\nonumber
\\
 + \; {\Gamma (a + b + 1 - c) \;\Gamma (c -1) \over \Gamma (a)
\; \Gamma (b) } \; {1 \over r^{j+1} } \; (1 - r^{2})^{-i\epsilon
/2}
 F (b - c + 1, a - c + 1,- c + 2; r^{2})  \;  ,
\nonumber
\\
\label{6.2}
\end{eqnarray}

\noindent where
\begin{eqnarray}
a =  { j - i \rho \epsilon   \over 2} \; , \;\;
b =  { j + 1 - i \epsilon  \rho  \over 2} \; , \; \;
c =    j + 3/2 = p + 1 \; .
\nonumber
\end{eqnarray}

\noindent At far  distance away from the~horizon ($R << \rho )$, for
the $U^{out}_{run}(R)$ one has  approximated representation
\begin{eqnarray}
U^{out}_{run}(R)  \sim   \left [ \;
{\Gamma (a + b + 1 - c)\; \Gamma (1 - c)  \over
\Gamma (b - c  + 1) \Gamma (a - c +  1)} \; {1 \over \rho ^{j} }\;
 ({2 \over \epsilon })^{p}
\; \Gamma (1 + p )\; {J_{p}(\epsilon R) \over \sqrt{R}}
\right.
\nonumber
\\
\left. +\; {\Gamma (a + b  + 1  - c)\; \Gamma (c - 1) \over \Gamma (a)
\Gamma (b)} \; \rho ^{j +1} \; ({2 \over \epsilon })^{-p} \;
\Gamma (1 - p )\; {J_{-p}(\epsilon R) \over \sqrt{R}} \; \right ] \; ,
\label{6.3a}
\end{eqnarray}

\noindent but one  ought to remember the concomitant  limitation
(\ref{4.18}) on quantum numbers
\begin{eqnarray}
 R << \rho   \;, \;\; j << \epsilon \rho  \; .
\label{6.3b}
\end{eqnarray}

\noindent This means that  to be correct, one has to take the
limits
\begin{eqnarray}
\lim_{ \epsilon \rho  >> j }  = \left [ \; {\Gamma (a + b + 1 - c)
\Gamma (1 - c)  \over \Gamma (b - c + 1) \Gamma (a - c + 1)}
\; \right ] \;\; ,
\qquad
\lim_{ \epsilon \rho >> j } = \left [\; {\Gamma (a  + b + 1 - c)
 \Gamma (c - 1) \over  \Gamma (a) \Gamma (b) } \; \right ].
 \label{6.3c}
\end{eqnarray}

\noindent
For the moment, setting away the need to
realize  additionally the~above limiting passages, let us proceed
with calculations (potentially) providing an  analytical
expression for $R_{\epsilon j}$.   So, taking into consideration  an~asymptotic formula for
Bessel functions
\begin{eqnarray}
x >> \nu ^{2}\;\;  : \;\;\;
J (x) \sim  {\Gamma (2 \nu + 1)\; 2^{-2\nu -1/2} \over
 \Gamma (\nu + 1) \; \Gamma (\nu + 1/2) } \; {1 \over \sqrt{x}}
\nonumber
\\
\times
\; \left [\; \exp \left ( + i (x  - {\pi \over 2} (\nu +{1\over 2}))
\right ) \; + \;
        \exp \left ( - i (x -  {\pi \over 2} (\nu + {1\over
2}))\right ) \; \right ] ,
\label{6.4}
\end{eqnarray}

\noindent we get
\begin{eqnarray}
j < j ^{2} << \epsilon R << \epsilon  \rho \; ,
\nonumber
\\
U^{out}_{run}(R) \sim
\left [ {e^{+i\epsilon R} \over \epsilon  R}
\left ( A \; \exp (- i {\pi \over 2} (p + {1 \over 2})) \; +\;
        B \; \exp (- i {\pi \over 2}(-p + {1 \over 2}) \right)
\right.
\nonumber
\\
\left.  +\; {e^{-i\epsilon R} \over \epsilon  R}
\left ( A \; \exp (+ i {\pi \over 2} (p + {1 \over 2})) \;+\;
        B \; \exp (+ i {\pi \over 2}(-p + {1 \over 2}))
\right ) \right ]    \; ,
\label{6.5a}
\end{eqnarray}

\noindent where $A$ and $B$ are
\begin{eqnarray}
A =  {\Gamma (a + b + 1 - c) \;\Gamma (1 - c) \over
      \Gamma (b - c + 1) \; \Gamma (a - c + 1)} \;
{2^{-j-1} \; \Gamma (2 p + 1 ) \over (\epsilon \rho)^{j}\;
\Gamma (p + 1/2)} \; ,
\nonumber
\\
B = {\Gamma (a + b + 1 - c) \; \Gamma (c - 1) \over \Gamma (a) \;
\Gamma (b) } \; {(\epsilon  \rho )^{j +1}\; \Gamma (- 2 p  + 1)
\over 2^{-j}\; \Gamma ( p  + 1/2 )} \; .
\label{6.5b}
\end{eqnarray}

\noindent
The reflection coefficient $R_{\epsilon j}$, by definition, is
a~square modulus of the~ratio of the~amplitude at
$ e^{-i\epsilon R} / \epsilon  R $   to the~amplitude at
$ e^{+i\epsilon R} / \epsilon R$ ; thus we  find
\begin{eqnarray}
R_{\epsilon j}  =  \left |
i \;e^{-i \pi p} \; {(A/B) \;e^{+i\pi p} \;+\; 1  \over
                (A/B)\; e^{-i\pi p}\; +\; 1 }   \right |  ^{2} \; .
\label{6.5c}
\end{eqnarray}

\noindent  It remains to show that, after taking into account
the~limitation $\epsilon \rho >> j$, this expression (\ref{6.5c}) for
$R_{\epsilon j}$  will  vanish identically.
For $A/B$ one has
\begin{eqnarray}
{A \over B} = {1 \over (2 \epsilon  \rho )^{2j+1} } \; {\Gamma (a)
\; \Gamma (b) \over \Gamma (a - c + 1)\; \Gamma (b - c + 1)} \;
\nonumber
\\
\times
{\Gamma (1 - c) \over \Gamma (c - 1)} \; {\Gamma (-p +1/2)  \over
\Gamma (p + 1/2)} \; {\Gamma ( 2p + 1)  \over \Gamma (- 2p + 1)}
\; . \label{6.6}
\end{eqnarray}

\noindent After using the formulas
\begin{eqnarray}
\Gamma (p) \; \Gamma (1  - p) =  { \pi  \over \sin (\pi p)} \;\; ,\qquad
{ \Gamma (2 x) \over \Gamma ( x )} = \Gamma (x + 1/2)\;
{2^{2x -1} \over \Gamma (1/2) }
\end{eqnarray}

\noindent we find
\begin{eqnarray}
{\Gamma (1 - c) \over \Gamma (c - 1)}\; {\Gamma (-p + 1/2)  \over
\Gamma (p + 1/2)}\; {\Gamma (2p + 1) \over  \Gamma (- 2p + 1)} = -
2^{4j + 2} \; .
\label{6.7}
\end{eqnarray}

\noindent Further, applying  the~asymptotic relation
\begin{eqnarray}
\{ \; x \rightarrow  \infty \; , \;  \mid \arg x \mid
< \pi\;  \} \; \rightarrow \;  {\Gamma (x + \alpha ) \over \Gamma (x
+ \beta )} \sim x^{\alpha -\beta }\; ,
\nonumber
\end{eqnarray}

\noindent we get
\begin{eqnarray}
\lim_{ \epsilon \rho >> j } = \left [  \; { \Gamma (a)\; \Gamma
(b) \over \Gamma (a - c + 1) \; \Gamma (b - c + 1) } \; \right ] =
({\epsilon p \over 2})^{2j+1} \; (- i )^{2p} \; .
\label{6.8}
\end{eqnarray}

\noindent Substituting   (\ref{6.7})  and (\ref{6.8}) into  (\ref{6.6}), for $A/B$
we finally arrive at  very simple result
\begin{eqnarray}
\lim_{\epsilon \rho  >> j } {A \over B} = - ( - i )^{2p}
\label{6.9}
\end{eqnarray}

\noindent and therefore the  identical zero  for the
$R_{\epsilon j}$ arises, $R_{\epsilon j}   \equiv 0 $.

In other words,
analytical expressions for $R_{\epsilon j} $  are due to the fact of not taking into account
 all required restrictions.

\section{Approximations  influencing  physical results}

To prevent anyone from possible errors
let us discuss the formula   (\ref{6.3a}) giving decomposition for a wave passing to de Sitter horizon
at far distance domain and consider possible discrepancies arising from
the use of incorrect approximate expressions for wave functions.
Indeed, instead of correct   decom\-position  (\ref{6.3a}) let us write its (by hand) modification
through introducing at  $J_{p}(\epsilon R)$ and
 $J_{-p}(\epsilon R)$  some coefficients
 $(1+\Delta_{p})$  and     $(1+\Delta_{-p})$  (here, the correct formula  corresponds to
$\Delta_{p}=0$  and     $\Delta_{-p}=0$):
\begin{eqnarray}
U_{out}(R)  \sim   [ \;
(1+\Delta_{p})
{\Gamma (a + b + 1 - c)  \Gamma (1 - c)  \over
\Gamma (b - c  + 1) \Gamma (a - c +  1)}  {1 \over \rho ^{j} }
 ({2 \over \epsilon })^{p}
\; \Gamma (1 + p )  {J_{p}(\epsilon R) \over \sqrt{R}}
\nonumber
\\
+ \;   (1+\Delta_{-p})
{\Gamma (a + b  + 1  - c)  \Gamma (c - 1) \over
\Gamma (a) \Gamma (b)}  \rho ^{j +1}
({2 \over \epsilon })^{-p}  \Gamma (1 - p )\; {J_{-p}(\epsilon R)
\over \sqrt{R}} \;  ]   \; .
\nonumber
\\
\label{7.1}
\end{eqnarray}

\noindent Correspondingly, instead of  $A$ and $B$ according to  (\ref{6.5b}) we have
\begin{eqnarray}
b
A \; \rightarrow \; A' = (1+\Delta_{p}) \; A \;\; , \;\;
B \; \rightarrow \; B' = (1+\Delta_{-p})\; B \; ,
\nonumber
\label{7.2}
\\
\lim_{\epsilon\rho >>j}\; {A \over B} \; \rightarrow \;
\lim_{\epsilon\rho >>j} \; {A' \over B'} = \lim_{\epsilon\rho >>j}
\; {1+\Delta_{p} \over 1+\Delta_{-p} } \; \left [ -(-i)^{2p} \right ] \; .
\label{7.3}
\end{eqnarray}

\noindent Therefore,   we arrive at a new expression for the
reflection coefficient  (compare it with (\ref{6.5c}))
\begin{eqnarray}
R_{\epsilon j}'  = \;  \left |  \;
{(A'/B') (+i)^{2p} + 1  \over   (A'/B') (-i)^{2p} + 1   } \; \mid^2 \; =
\mid\; {\Delta_{p} - \Delta_{-p} \over 2 + \Delta_{p} + \Delta_{-p}}\; \right |  ^2 \; .
\label{7.4}
\end{eqnarray}

\noindent In other words, just an error  in approximate form for
a wave  function leads us to a  "physical result".

\section{ On Maxwell equations in  general covariant tetrad  Majorana -- Oppenheimer  form
on the background of  de Sitter  space-time
 }

Below we show that the main results can be extended to the case of electro\-magnetic field.
To this end, we will use an old and almost unusable in the literature  approach
 by Riemann -- Silberstein -- Majorana -- Oppenheimer
in general covariant tetrad form.
Maxwell equations in Riemann space can be presented  as one matrix equation
\cite{Red'kov-Bogush-Tokarevskaya-Spix-2009},\cite{Bogush-Krylov-Ovsiyuk-Red'kov},
\cite{Book}
\begin{eqnarray}
\alpha^{1} = \left | \begin{array}{rrrr}
0 & 1  &  0  & 0  \\
-1 & 0  &  0  & 0  \\
0 & 0  &  0  & -1  \\
0 & 0  &  1  & 0
\end{array}  \right |  , \alpha^{2} = \left | \begin{array}{rrrr}
0 & 0  &  1  & 0  \\
0 & 0  &  0  & 1  \\
-1 & 0  &  0  & 0  \\
0 & -1  & 0  & 0
\end{array}  \right | ,
\nonumber
\\
\alpha^{3} = \left | \begin{array}{rrrr}
0 & 0  &  0  & 1  \\
0 & 0  & -1  & 0  \\
0 & 1  &  0  & 0  \\
-1 & 0  &  0  & 0
\end{array}  \right |, \qquad  \alpha^{0} = -i I\; ,
\nonumber
\\
 \Psi = \left | \begin{array}{c} 0 \\ {\bf E} + i c{\bf B}
\end{array} \right | \; , \qquad J
= {1 \over \epsilon_{0}} \; \left | \begin{array}{c} \rho \\
i{\bf j}
\end{array} \right |,
\nonumber
\\
\alpha^{c} \; ( \; e_{(c)}^{\rho} \partial_{\rho} + {1 \over 2}
j^{ab} \gamma_{abc} \; ) \; \Psi = J(x)\; , \label{3.1'}
\end{eqnarray}

\noindent  or
\begin{eqnarray}
-i  (  e_{(0)}^{\rho} \partial_{\rho} + {1 \over 2} j^{ab}
\gamma_{ab0}  )\Psi + \alpha^{k}  (  e_{(k)}^{\rho}
\partial_{\rho} + {1 \over 2} j^{ab} \gamma_{abk}  )\Psi =
J(x)\; . \label{3.2'}
\end{eqnarray}

\noindent Allowing for identities
\begin{eqnarray}
{1 \over 2} j^{ab} \gamma_{ab0} = [ s_{1} ( \gamma_{230} +i
\gamma_{010} ) + s_{2} ( \gamma_{310} +i \gamma_{020}) + s_{3} (
\gamma_{120} +i \gamma_{030} ) ] \; , \nonumber
\\
{1 \over 2} j^{ab} \gamma_{abk} = [ s_{1} ( \gamma_{23k} +i
\gamma_{01k} ) + s_{2} ( \gamma_{31k} +i \gamma_{02k}) + s_{3} (
\gamma_{12k} +i \gamma_{03k} ) ] \; ,
\nonumber \label{3.3'}
\end{eqnarray}

\noindent where
\begin{eqnarray}
s_{1}=   \left | \begin{array}{cc} 0 & 0 \\0 & \tau_{1} \end{array}
\right |,
 s_{2} =  \left | \begin{array}{cc}
0 & 0 \\0 & \tau_{1} \end{array} \right |  , \;
s_{3} =  \left | \begin{array}{cc}
0 & 0 \\0 & \tau_{1} \end{array} \right |  ,
\nonumber
\\
\tau_{1}= \left | \begin{array}{rrr}
 0 & 0 & 0 \\
 0 & 0 & -1 \\
 0 & 1 & 0 \\
\end{array} \right |
, \;
 \tau_{2} = \left | \begin{array}{rrr}
0 & 0 & 1 \\
0 & 0 & 0 \\
-1 & 0 & 0 \\
\end{array} \right | ,\;
\tau_{3} = \left | \begin{array}{rrr}
0 & -1 & 0 \\
1 & 0 & 0 \\
0 & 0 & 0
\end{array} \right |  ,
\label{tau}
\end{eqnarray}

\noindent and using the notation
\begin{eqnarray}
e_{(0)}^{\rho} \partial_{\rho} = \partial_{(0)} \; , \qquad
e_{(k)}^{\rho} \partial_{\rho} = \partial_{(k)} \; , \qquad
a =0,1,2,3 \; , \nonumber
\\
( \gamma_{01a}, \gamma_{02a} , \gamma_{03a} ) = {\bf v}_{a} \; ,
\qquad ( \gamma_{23a}, \gamma_{31a} , \gamma_{12a} ) = {\bf p}_{a}
\; , \label{3.4'}
\end{eqnarray}

\noindent eq. (\ref{3.2'})  in the absence of sources reduces to
\begin{eqnarray}
-i  [ \;  \partial_{(0)} + {\bf s} ({\bf p}_{0} +i{\bf v}_{0}
) \; ] \Psi    + \alpha^{k}\;   [ \; \partial_{(k)} + {\bf s} ({\bf
p}_{k} +i{\bf v}_{k}  )\;  ] \;   \Psi =  0 \; .
\label{3.5'}
\end{eqnarray}

Let us consider this  equation  in   de Sitter static   metrics and tetrad
\begin{eqnarray}
 dS^{2} = \Phi \;  dt^{2} - {dr^{2} \over \Phi } -
r^{2} (d\theta^{2} +  \sin ^{2}\theta d\phi ^{2}) \; ,  \; \; \Phi  =  1 - r^{2}  \; ,
\nonumber
\\
e^{\alpha}_{(0)}=({1 \over  \sqrt{\Phi} }, 0, 0, 0) \; , \qquad
e^{\alpha}_{(3)}=(0, \sqrt{\Phi}, 0, 0) \; , \nonumber
\\
e^{\alpha}_{(1)}=(0, 0, \frac {1}{r}, 0) \; , \qquad
e^{\alpha}_{(2)}=(1, 0, 0, \frac{1}{ r  \sin \theta})  \; ,
\nonumber
\\
\gamma_{030} ={ \Phi ' \over 2\sqrt{ \Phi }} \; , \; \gamma_{311} ={
\sqrt{ \Phi } \over r} \; , \; \gamma_{322} ={ \sqrt{ \Phi }
\over r} \; , \; \gamma_{122} ={ \cos \theta \over r \sin
\theta} \; ,
\end{eqnarray}

\noindent we get to explicit form of the matrix
equation
\begin{eqnarray}
 [   -  { i  \partial_{t} \over \sqrt{ \Phi }}  + \sqrt{\Phi}  (
 \alpha^{3}  \partial_{r}   + {  \alpha^{1}  s_{2}  - \alpha^{2}  s_{1}   \over r} +
    { \Phi'   \over  2 \Phi }  s_{3}  )
   +  {1 \over r }   \Sigma_{\theta, \phi}       ]
 \left | \begin{array}{c}
0 \\ \psi
\end{array} \right | = 0 \; ,
\nonumber
\\
\Sigma_{\theta, \phi} =
  { \alpha^{1}  \over r}\partial_{\theta} +
 \alpha^{2} \;  { \partial_{\phi} +   s_{3}   \cos \theta  \over  \sin \theta } \;   .
\label{4.11'}
\end{eqnarray}

\noindent
It is convenient to have the   $s_{3}$ as  a diagonal matrix, that is reached by a simple
linear transformation to the known cyclic basis
\begin{eqnarray}
 \Psi ' = U_{4} \Psi  \; , \qquad
\; U_{4}  = \left | \begin{array}{cc} 1 & 0 \\
0 &   U
\end{array} \right |  ,
\nonumber
\\
U = \left |
 \begin{array}{ccc}
- 1 /\sqrt{2}  &\;  i /  \sqrt{2}  &\;  0  \\
0  &  \; 0  & \;  1  \\
1 / \sqrt{2}  &\;   i  / \sqrt{2}  & \;  0
\end{array} \right |  , \;
U^{-1}  = \left |
 \begin{array}{ccc}
- 1 /\sqrt{2}  & \;  0  & \; 1 /  \sqrt{2}    \\
-i / \sqrt{2}  &\;   0  & \;  -i / \sqrt{2}   \\
0    &  \; 1    & \;  0
\end{array} \right |  ,
\label{cyclic}
\end{eqnarray}

\noindent  so that
\begin{eqnarray}
  \tau'_{1}  =  {1 \over \sqrt{2}} \left |
\begin{array}{ccc}
0  &  -i   &  0  \\
-i  &  0  &  -i  \\
0  &  -i  &  0
\end{array} \right |  ,\qquad
  \tau'_{2}  = {1 \over \sqrt{2}} \left |
\begin{array}{ccc}
0  &  -1  &  0  \\
1  & 0  &  -1  \\
0  &  1  &  0
\end{array} \right |  ,\;\;
\nonumber
\end{eqnarray}
\begin{eqnarray}
 \alpha^{'1} = {1 \over \sqrt{2}}
\left | \begin{array}{rrrr}
0 & - 1 & 0 & 1 \\
1 & 0 & -i & 0 \\
0 & -i & 0 & - i \\
-1 & 0 & -i & 0
\end{array} \right |  ,
\alpha^{'2} = {1 \over \sqrt{2}} \left | \begin{array}{rrrr}
0 & - i & 0 & -i \\
-i & 0 & -1 & 0 \\
0 & 1 & 0 &  -1 \\
-i & 0 & 1 & 0
\end{array} \right |  ,
\nonumber
\\
 \tau'_{3} = - i \; \left | \begin{array}{rrr}
+1  &  0  &  0  \\
0  &  0  &  0   \\
0  &  0  &  -1
\end{array} \right | , \qquad
  \alpha^{'3}  =  \left
| \begin{array}{rrrr}
0  & 0  & 1 & 0 \\
0  & -i & 0 & 0 \\
-1 &  0 & 0 & 0 \\
0  &  0 & 0 & +i
\end{array} \right |  .
\nonumber
\end{eqnarray}

Eq.  (\ref{4.11'})  becomes
\begin{eqnarray}
 [   -  { i  \partial_{t} \over \sqrt{ \Phi }}  + \sqrt{\Phi}  (
 \alpha^{'3}  \partial_{r}   + {  \alpha^{'1}  s_{'2}  - \alpha^{'2}  s'_{1}   \over r} +
    { \Phi ' \over   2 \Phi } s'_{3}  )
   +  {1 \over r }   \Sigma\;'_{\theta, \phi}       ]
 \left | \begin{array}{c}
0 \\ \psi'
\end{array} \right | = 0 \; ,
\nonumber
\\
\Sigma\;'_{\theta, \phi} =
  { \alpha^{'1}  \over r}\partial_{\theta} +
 \alpha^{'2} \;  { \partial_{\phi} +   s'_{3}   \cos \theta  \over  \sin \theta } \;   .
\label{4.11''}
\end{eqnarray}

\section{Separation of variables and Wigner functions}

\hspace{5mm}
 Spherical  waves with $(j,m)$ quantum numbers
should be constructed as follows
\begin{eqnarray}
\psi  = e^{-i \omega t} \left | \begin{array}{c}
0 \\
f_{1}(r) D_{-1 }
\\
f_{2}(r) D_{0 }  \\
f_{3}(r) D_{+1 }
\end{array} \right |
\label{5.1'}
\end{eqnarray}

\noindent where the  notation for Wigner $D$-functions
\cite{Wigner-1927}, \cite{Varshalovich-Moskalev-Hersonskiy-1975} is used
\begin{eqnarray}
 D_{\sigma} = D^{j}_{-m, \sigma} ( \phi , \theta, 0)\;, \; \sigma
= -1, 0, +1 \; ;
\nonumber
\end{eqnarray}

\noindent
 $j,m$  determine total angular momentum. We adhere
notation developed in
 \cite{Red'kov-1998(3)}, \cite{Red'kov-1998(4)}, \cite{Red'kov-1999(1)},
 \cite{Red'kov-1999(2)}; earlier similar  techniques  was used
by Dray \cite{Dray-1985}, \cite{Dray-1986},
Krolikowski and Turski
\cite{Krolikowski-Turski-1986},
Turski \cite{Turski-1986};  many years ago such a tetrad basis was  used by
 Schr\"{o}dinger \cite{Schrodinger-1938}
 and Pauli \cite{Pauli-1939} when considering the problem
of single-valuedness of wave  functions in quantum theory.
In the literature the equivalent techniques of  spin-weighted harmonics
Goldberg et al  \cite{Goldberg-Macfarlane-Newman-Rohrlich-Sudarshan-1967} (se also in
\cite{Penrose-Rindler-1984})
is used preferably though the equivalence of both approaches is known \cite{Dray-1985}, \cite{Dray-1986}.

With the use  of the  recursive relations  \cite{Varshalovich-Moskalev-Hersonskiy-1975}
(below $\nu =  \sqrt{j(j+1)} \; , \;  a = \sqrt{(j-1)(j+2)}\;$)
\begin{eqnarray}
\partial_{\theta}   D_{-1} =   {1 \over 2}  (
a   D_{-2} -    \nu   D_{0}  ) \; , \; \frac {m -
\cos{\theta}}{\sin {\theta}}  D_{-1} = {1 \over 2} (  a
D_{-2} + \nu  D_{0}  ) \; , \nonumber
\\
\partial_{\theta}   D_{0} = {1 \over 2}  (
\nu   D_{-1} - \nu   D_{+1}  ) \; , \qquad  \frac {m}{\sin
{\theta}}  D_{0} = {1 \over 2}  (    \nu  D_{-1} +  \nu
D_{+1}  ) \; , \nonumber
\\
\partial_{\theta}  D_{+1} =
{1 \over 2}  (  \nu   D_{0} - a   D_{+2}  ) \; , \;
\frac {m + \cos{\theta}}{\sin {\theta}}   D_{+1} ={1 \over 2} \;
(
 \nu  D_{0} + \ a   D_{+2}  ) \;  ,
\label{5.2'}
\end{eqnarray}

\noindent we get
 (the factor  $e^{-i \omega
t} $ is omitted)
\begin{eqnarray}
 \Sigma\;'_{\theta \phi} \Psi ' =
 { \nu  \over \sqrt {2}} \;
 \left | \begin{array}{r}
 (f_{1} +f_{3}) D_{0} \\
 -i \;  f_{2} D_{-1}
 \\
 i \;  (f_{1} -f_{3}) D_{0}
 \\
 + i \;  f_{2} D_{+1}
 \end{array} \right  |.
\label{5.3'}
\end{eqnarray}

\noindent
Turning back to the Maxwell equation  (\ref{4.11''}),  we arrive at
the radial system
\begin{eqnarray}
1) \qquad   \sqrt{ \Phi }\; ( {d \over d r} +  {2   \over r }
)\;f_{2} + {1 \over r } \;
 { \nu  \over \sqrt {2}} \;  (f_{1} +f_{3})= 0\; ,
\nonumber
\\
2) \qquad   ( - {\omega \over \sqrt{ \Phi }}   - i\; \sqrt{
\Phi }\; {d \over dr } - i {\sqrt{ \Phi }  \over r }  - i{ \Phi '\over
2 \sqrt{ \Phi }}  ) \;f_{1} -
 {i \over r } \;  { \nu  \over \sqrt {2}} \; f_{2} = 0 \; ,
\nonumber
\\
3) \qquad  \qquad - {\omega \over \sqrt{ \Phi }} f_{2} + {i \over
r } \;
 { \nu  \over \sqrt {2}} (f_{1} -f_{3}) = 0 \; ,
\nonumber
\\
4) \qquad  (- {\omega \over \sqrt{ \Phi }} + i \; \sqrt{
\Phi }\;{d \over dr }  + i {\sqrt{ \Phi }  \over r } +i { \Phi '\over
2 \sqrt{ \Phi }} ) \;f_{3} + {i \over r } \;  { \nu  \over
\sqrt {2}} \; f_{2} = 0 \; . \label{5.5'}
\end{eqnarray}

\noindent
Combining equations  2) and  4),  instead of  (\ref{5.5'}) we get

\vspace{2mm}

$
 2) + 4)\; , \;
$
\begin{eqnarray}
  - {\omega \over \sqrt{ \Phi }} (f_{1} + f_{3})   -
i   ( \sqrt{ \Phi } {d \over dr } + {\sqrt{ \Phi } \over
r } + { \Phi '\over 2 \sqrt{ \Phi }}   )   (f_{1} - f_{3}) = 0 \;
, \nonumber
\end{eqnarray}

$2) -4) \; , \;
 $
 \begin{eqnarray}
  - {\omega \over \sqrt{ \Phi }}   (f_{1} - f_{3})  - i (  \sqrt{ \Phi
} {d \over dr }  + {\sqrt{ \Phi }  \over r } + { \Phi '\over 2  \sqrt{
\Phi }}   ) (f_{1} + f_{3})  -
 {2i \over r }   { \nu  \over \sqrt {2}}  f_{2}  = 0 \; ,
\nonumber
\\
3) \qquad  - {\omega \over \sqrt{ \Phi }} f_{2} + {i \over
r }  { \nu  \over \sqrt {2}} (f_{1} -f_{3}) = 0 \; ,
\nonumber
\\
1) \qquad   \sqrt{ \Phi } ( {d \over d r} +  {2   \over r }
)f_{2} + {1 \over r }
 { \nu  \over \sqrt {2}}   (f_{1} +f_{3})= 0\; .
\nonumber
\label{3.9}
\end{eqnarray}

\noindent
It is easily verified that equation  1) is an identity when accounting for remaining ones.
So independent equations are
\begin{eqnarray}
- {\omega \over \sqrt{ \Phi }} f_{2} + {i \over r } \;
 { \nu  \over \sqrt {2}} (f_{1} -f_{3}) = 0 \; ,
\nonumber
\\
  - {\omega \over \sqrt{ \Phi }} (f_{1} + f_{3})   -
i\;   ( \sqrt{ \Phi }\; {d \over dr } + {\sqrt{ \Phi } \over
r } + { \Phi '\over  2 \sqrt{ \Phi }}   )   (f_{1} - f_{3}) = 0 \;
, \nonumber
\\
  - {\omega \over \sqrt{ \Phi }}   (f_{1} - f_{3})  - i\; (  \sqrt{ \Phi
} {d \over dr }  + {\sqrt{ \Phi }  \over r } + { \Phi '\over  2 \sqrt{
\Phi }}  )  (f_{1} + f_{3})  -
 {2i \over r }   { \nu  \over \sqrt {2}}  f_{2}  = 0 \; .
\nonumber
\\
\label{3.10'}
\end{eqnarray}

Let us introduce new functions
\begin{eqnarray}
f = { f_{1} + f_{3}  \over \sqrt{2}}   \; , \qquad g = { f_{1}-
f_{3}  \over \sqrt{2}}   \; , \nonumber
\end{eqnarray}

\noindent then eqs.  (\ref{3.10'}) read
\begin{eqnarray}
 f_{2} = {i  \nu \over \omega  }   { \sqrt{ \Phi }  \over r }  g   \; ,
\;\; - {\omega \over  \Phi }  f   -
i    (  {d \over dr } + {1  \over r } + { \Phi ' \over  2 \Phi }
 )   g = 0 \; , \nonumber
\\
  - {\omega^{2} \over \Phi  }   g  - i \omega   (   {d \over dr }  + { 1  \over r } + { \Phi '\over
2 \Phi }  ) \; f  +
 {\nu^{2}  \over r^{2} } \;    \; g   = 0 \; .
\label{3.12'}
\end{eqnarray}

The system   (\ref{3.12'})  is simplified by substitutions
\begin{eqnarray}
g (r) = {1\over r \sqrt {\Phi }}\; G (r) \; , \qquad  f(r)  = {1\over r \sqrt
{\Phi}}\;  F (r) \; ,
\nonumber
\end{eqnarray}

\noindent and it gives
\begin{eqnarray}
 f_{2} = {i  \nu \over  \omega  } \;
 {1\over r^{2}  }\; G (r)   \; , \qquad
 i  \omega \;  F  = \Phi {d \over dr } G \; ,
\nonumber
\\
  i \omega \;   {d \over dr } \; F  +   {\omega^{2} \over \Phi  }
    \; G   - {\nu^{2} \over r^{2}} \; G   = 0 \; .
\label{3.13'}
\end{eqnarray}

\noindent So we arrive  at a    differential equation
for  $G (r)$
\begin{eqnarray}
{d^{2}G\over d r^{2}}+{\Phi'\over \Phi}\;{d G \over
dr}+\left({\omega^{2}\over \Phi^{2}}-{j(j+1)\over
r^{2}\Phi}\right)G=0 \; ,
\nonumber
\label{3.14}
\end{eqnarray}

\noindent
or taking $\Phi = 1 - r^{2}$
\begin{eqnarray}
{d^{2}G\over d r^{2}} -{2r \over 1 -r^{2} }\;{d G \over
dr}+\left({\omega^{2}\over (1-r^{2})^{2}}-{j(j+1)\over
r^{2} (1-r^{2}) }\right) G=0 \; . \label{3.14a}
\end{eqnarray}

This equation coincides with that followed  from (\ref{2.3}), if   one translates
eq.  (\ref{2.3})  at $M=0$  to a new function
$
f(r) =  r^{-1} \varphi (r) $:
\begin{eqnarray}
{d^{2} \varphi  \over d r^{2}} -  {2r \over 1 -r^{2} }\;
{d \varphi  \over dr}+
\left ( {\omega^{2}\over (1-r^{2})^{2}}-{j(j+1)\over
r^{2} (1-r^{2}) } \right )  \varphi = 0\;.
\label{3.14b}
\end{eqnarray}

With the variable  $z= r^{2}$, eq. (\ref{3.14b}) gives
\begin{eqnarray}
4z(1-z){d^{2}G\over dz^{2}}+2(1-3z){dG\over dz}+\left({\omega^{2}\over 1-z}-{ j(j+1) \over
z}\right)G=0 \; ,
\label{1.2}
\end{eqnarray}

\noindent and after substitution $
G = z^{a} (1-z)^{b} F (z)$ we arrive at
\begin{eqnarray}
4z(1-z) {d^{2}F\over dz^{2}} + 4\left[2a + {1\over2}-(2a+2b+{3\over2})z\right]{dF\over dz}
\nonumber
\\
+\left[{4a^{2}-2a- j(j+1)\over z} +{4b^{2}+\omega^{2}\over 1-z}-4(a+b)(a+b+{1\over2})\right]F=0.
\label{1.3}
\end{eqnarray}

\noindent
Requiring
\begin{eqnarray}
 a=  + {j+1 \over 2} \;, \;  -{j \over 2} \;  , \qquad  b = \pm \;
{  i\omega  \over 2} \; , \; \omega > 0 \;  ;
\nonumber
\end{eqnarray}

\noindent
we get
\begin{eqnarray}
z(1-z){d^{2}F\over dz^{2}}
+
\left [ 2a+{1\over2}-(2a+2b+{3\over2})z \right ]
{dF\over dz}
\nonumber
\\
-(a+b)(a+b+{1\over2}) \; F=0 \; ,
\label{1.5}
\end{eqnarray}

\noindent where the parameters of the hypergeometric function are given by
\begin{eqnarray}
\alpha = a+b \;, \qquad \beta = a+b+ {1\over2} \;, \qquad \gamma = 2a +{1\over2} \; .
 \label{1.6}
\end{eqnarray}

\noindent
Evidently, these  solutions coincide with those described in Sec. {\bf 2}.

\section{Spin 1/2 particle in de Sitter space }

The Dirac equation (the notation according \cite{Book-2009} is used)
\begin{eqnarray}
[\; i \gamma^{c} \; ( e_{(c)}^{\alpha} \partial_{\alpha} + {1 \over 2} \sigma^{ab} \gamma_{abc}) - M \; ] \; \Psi = 0
\label{Dirac}
\end{eqnarray}

\noindent
in static coordinates and tetrad of the Sitter space takes the form
\begin{eqnarray}
[  \;i\; {\gamma ^{0}  \over \sqrt{\Phi}} \partial _{t}  + i \sqrt{\Phi } (     \gamma ^{3}
\partial _{r}  + { \gamma ^{1} \sigma^{31}  +
\gamma  ^{2} j^{32} \over r }   +
  { \Phi' \over 2 \Phi  } \gamma ^{0}     \sigma^{03} )
\nonumber
\\
+  {
1 \over r} \Sigma _{\theta,\phi }  -  M\;
  ]   \; \Psi (x)  =  0 \; ,\qquad
\Sigma _{\theta ,\phi } =  \; i\; \gamma ^{1}
\partial _{\theta } + \gamma ^{2} {i\partial +i\sigma^{12}\cos
\theta  \over \sin \theta  } \; .
\nonumber
\\
\label{10.1a}
\end{eqnarray}

\noindent
Below the spinor basis will be used
\begin{eqnarray}
\gamma^{0} =
\left | \begin{array}{cc}
0 & I \\
I & 0
\end{array} \right |, \qquad   \gamma^{j} =
\left | \begin{array}{cc}
0 & -\sigma_{j} \\
\sigma_{j}  & 0
\end{array} \right |, \qquad   i \sigma^{12} =
\left | \begin{array}{cc}
\sigma_{3} & 0 \\
0 & \sigma_{3}
\end{array} \right | .
\nonumber
\end{eqnarray}

\noindent
Allowing for $
\gamma ^{1} \sigma^{31}  +
\gamma  ^{2} j^{32} = \gamma^{3}, \;
\gamma ^{0} \sigma^{03}  = \gamma^{3}/  2$, eq. (\ref{10.1a}) reads
\begin{eqnarray}
[ \;i\; {\gamma ^{0}  \over \sqrt{\Phi} }   \partial _{t} \; + i \sqrt{\Phi } \gamma^{3}\; (  \;
\partial _{r}  +   { 1 \over r }   +
  { \Phi' \over 4 \Phi  }  \;  )
+   { 1 \over r} \;\Sigma _{\theta,\phi } -  M \;
  ]  \; \Psi (x)  =  0  \; .
\label{10.1b}
\end{eqnarray}

\noindent
One can simplify  the problem with the help of substitution
$$
 \Psi  (x)  = r ^{-1} \Phi ^{-1/4} \; F (x)
 $$

 \noindent then
\begin{eqnarray}
[ \;i\; {\gamma ^{0}  \over \sqrt{\Phi} }   \partial _{t} \; + i \sqrt{\Phi } \gamma^{3}\;   \;
\partial _{r}  +   { 1 \over r} \;\Sigma _{\theta,\phi }  -  M \;
  ]  \; F (x)  =  0  \; .
\label{10.1c}
\end{eqnarray}

\noindent Spherical waves are constructed through the substitution
\begin{eqnarray}
\Psi _{\epsilon jm}(x) \;  = \; {e^{-i\epsilon t} \over r} \;
\left | \begin{array}{l}
        f_{1}(r) \; D_{-1/2} \\ f_{2}(r) \; D_{+1/2}  \\
        f_{3}(r) \; D_{-1/2} \\ f_{4}(r) \; D_{+1/2}
\end{array} \right | \; .
\label{10.2}
\end{eqnarray}

\noindent With the use of the recursive  relations \cite{Varshalovich-Moskalev-Hersonskiy-1975}
\begin{eqnarray}
\partial_{\theta} \; D_{+1/2} \; = \;  a\; D_{-1/2}  - b \; D_{+3/2}  \; ,
\nonumber
\\
{- m - 1/2 \;  \cos \theta  \over  \sin \theta } \; D_{+1/2}\; =\;
- a \;  D_{-1/2} - b \;  D_{+3/2}  \; ,
\nonumber
\\
\partial_{\theta} \;  D_{-1/2} \; = \;  b \; D_{-3/2} - a \; D_{+1/2}  \; ,
\nonumber
\\
{- m + 1/2 \; \cos \theta \over \sin \theta}  \;  D_{-1/2} \; = \;
 - b \; D_{-3/2}  - a \;  D_{+1/2}  \; ,
\nonumber
\end{eqnarray}

\noindent where
$
a = (j + 1)/2\;, \; b  = (1 / 2) \;\sqrt{(j-1/2)(j+3/2)}\;,
$
 we get
\begin{eqnarray}
 \Sigma _{\theta ,\phi } \; \Psi _{\epsilon jm}(x) \;  = \; i\; \nu \;
{e^{-i\epsilon t } \over r} \; \left | \begin{array}{r}
        - \; f_{4}(r) \; D_{-1/2}  \\  + \; f_{3}(r) \; D_{+1/2} \\
        + \; f_{2}(r) \; D_{-1/2}  \\  - \; f_{1}(r) \; D_{+1/2}
\end{array} \right |    , \;\; \nu  = (j + 1/2) \; .
\label{10.3}
\end{eqnarray}

\noindent  Then we arrive at the radial system
\begin{eqnarray}
{\epsilon  \over \sqrt{\Phi}}   f_{3}   -  i  \sqrt{\Phi} {d \over dr}  f_{3}   - i {\nu \over r}
f_{4}  -  M  f_{1} =   0  \; ,
\nonumber
\\
{\epsilon  \over \sqrt{\Phi}}   f_{4}   +  i \sqrt{\Phi}  {d
\over dr} f_{4}   + i {\nu \over r} f_{3}  -  M  f_{2} =   0    \;
,
\nonumber
\\
{\epsilon  \over \sqrt{\Phi}}   f_{1}   +  i \sqrt{\Phi}  {d \over dr}  f_{1}  + i {\nu \over r}
f_{2}  -  M  f_{3} =   0  \; ,
\nonumber
\\
{\epsilon   \over \sqrt{\Phi}} f_{2}   -  i\sqrt{\Phi}   {d
\over dr} f_{2}   - i {\nu \over r} f_{1}  -  M  f_{4} =   0  \; .
\label{10.4}
\end{eqnarray}

\noindent To simplify the system, let us diagonalize $P$-operator.
In Cartesian basis,  $\hat{\Pi}_{C.}  =  i \gamma ^{0} \otimes \hat{P}$,
after transition to spherical  tetrad gives
\begin{eqnarray}
\hat{\Pi}_{sph.} \; \; = \left | \begin{array}{cccc}
0 &  0 &  0 & -1   \\
0 &  0 & -1 &  0   \\
0 &  -1&  0 &  0   \\
-1&  0 &  0 &  0
\end{array} \right |
\; \otimes  \; \hat{P} \; .
\label{10.5}
\end{eqnarray}

\noindent From the equation
$\hat{\Pi}_{sph.} \; \Psi _{jm} = \; \Pi \; \Psi _{jm}$ it follows that
$\Pi =  \delta \;  (-1)^{j+1} , \; \delta  = \pm 1 $ and
\begin{eqnarray}
f_{4} = \; \delta \;  f_{1} , \;  f_{3} = \;\delta \; f_{2} \;
,\;
\Psi (x)_{\epsilon jm\delta } \; = \; {e^{-i\epsilon t} \over  r }
\; \left | \begin{array}{r}
     f_{1}(r) \; D_{-1/2} \\
     f_{2}(r) \; D_{+1/2} \\
\delta \; f_{2}(r) \; D_{-1/2}   \\
\delta \; f_{1}(r) \; D_{+1/2}
\end{array} \right |  \; .
\label{10.10}
\end{eqnarray}

\noindent Allowing for  (\ref{10.10}), we simplify the system  (\ref{10.4}) as
\begin{eqnarray}
( \sqrt{\Phi} {d \over dr} \;+\; {\nu \over r}\;) \; f \; + \; ( { \epsilon  \over \sqrt{\Phi}}  \;+
\;
 \delta \; M )\; g \; = \;0 \; ,
 \nonumber
\\
( \sqrt{\Phi} {d \over dr} \; - \;{\nu \over r}\;)\; g  \;- \; ( {\epsilon  \over \sqrt{\Phi}}\; -
\;
 \delta\;  M )\; f\; =\; 0     \; ;
\label{10.11}
\end{eqnarray}

\noindent where  new functions
\begin{eqnarray}
f \; = \; {f_{1} + f_{2} \over \sqrt{2}} \; , \qquad g \; = \;
{f_{1} - f_{2} \over i \sqrt{2}} \;
\nonumber
\end{eqnarray}
are used instead of   $f_{1}$  and  $f_{2}$.

For definiteness, let us consider eqs. (\ref{10.11}) at  $\delta = +1$
(formally the second case $\delta =-1$ corresponds to the
 change  $M \Longrightarrow - M$):
\begin{eqnarray}
( \sqrt{\Phi} {d \over dr} \;+\; {\nu \over r}\;) \; f \; + \; ( { \epsilon  \over \sqrt{\Phi}}  \;+
\;  M )\; g \; = \;0 \; ,
 \nonumber
\\
( \sqrt{\Phi} {d \over dr} \; - \;{\nu \over r}\;)\; g  \;- \; ( {\epsilon  \over \sqrt{\Phi}}\; -
\;
  M )\; f\; =\; 0     \; .
\label{10.12}
\end{eqnarray}

\noindent Here  there arise additional singularities at the points
$$
 \epsilon +
\sqrt{\Phi} \;  M  =0  \; , \qquad  \epsilon - \sqrt{\Phi} \;   M  = 0 \;.
$$
Correspondingly, the equation for $f(r)$ has  the form
\begin{eqnarray}
{d^{2}  \over dr^{2}} f -
\left ( {2r \over 1 -r^{2} } - {M r \over  \sqrt{1-r^{2}} (\epsilon + M \sqrt{1 -r^{2}})} \right )
{d \over dr } f
+ \left (
{ \epsilon^{2} \over (1-r^{2})^{2} } - {M^{2} \over 1 -r^{2}} \right.
\nonumber
\\
\left. -{ \nu (\nu +1) \over r^{2} (1 - r^{2} ) } -
{\nu \over (1-r^{2})   \sqrt{1-r^{2} } } +
 { M \nu  \over  \sqrt{1-r^{2}}   (\epsilon + M \sqrt{1 -r^{2}}) }
\right ) f = 0 \; .
\nonumber
\end{eqnarray}

However, there exists possibility to move these singularities away through a special
transformation of the functions $f(r), g(r)$ (see \cite{Otchik-1985}).
To this end, as a first step, let us  introduce a new variable $
r = \sin \rho  $, eqs. (\ref{10.12}) looks simpler
\begin{eqnarray}
( {d \over d \rho} \;+\; {\nu \over \sin \rho}\;) \; f \; + \; ( { \epsilon  \over \cos \rho }  \;+
\;  M )\; g \; = \;0 \; ,
 \nonumber
\\
(  {d \over d \rho} \; - \;{\nu \over \sin \rho}\;)\; g  \;- \; ( {\epsilon  \over \cos \rho }\; -
\;
 M )\; f\; =\; 0     \; .
\label{10.13}
\end{eqnarray}

\noindent Summing and subtracting two  last equations, we get
\begin{eqnarray}
{d \over d \rho} (f+g) + {\nu \over \sin \rho} (f-g) -  {\epsilon \over \cos \rho} (f-g) + M (f+g) = 0 \; ,
\nonumber
\\
{d \over d \rho} (f-g) + {\nu \over \sin \rho} (f+g) +  {\epsilon \over \cos \rho} (f+g) - M(f-g) = 0 \; .
\label{10.14}
\end{eqnarray}

\noindent Introducing two  new  functions
\begin{eqnarray}
f + g = e^{-i\rho/2} (F + G) \; , \qquad
f - g = e^{+i\rho/2} (F - G) \; ,
\label{10.15}
\end{eqnarray}

\noindent
one transforms (\ref{10.14})  into
\begin{eqnarray}
{d \over d \rho} e^{-i\rho/2} (F + G)  + {\nu \over \sin \rho} e^{+i\rho/2} (F - G)
\nonumber
\\
-  {\epsilon \over \cos \rho}  e^{+i\rho /2} (F - G)  + M e^{-i \rho /2} (F + G)  = 0 \; ,
\nonumber
\\
{d \over d \rho} e^{+i\rho /2} (F - G)  + {\nu \over \sin \rho} e^{-i\rho /2} (F + G)
\nonumber
\\
+  {\epsilon \over \cos \rho} e^{-i \rho /2} (F + G)  - M e^{+i \rho /2} (F - G)  = 0 \; ,
\nonumber
\end{eqnarray}

\noindent or
\begin{eqnarray}
 {d \over d \rho}  (F + G)  - {i \over 2} (F + G) +
{\nu \over \sin \rho} (\cos \rho + i \sin \rho)  (F - G)
 \nonumber
\\
-  {\epsilon \over \cos \rho}  (\cos \rho + i \sin \rho)  (F - G)  + M  (F + G)  = 0 \; ,
\nonumber
\end{eqnarray}
\begin{eqnarray}
{d \over d \rho}  (F - G)  + {i \over 2}  (F - G)  + {\nu \over \sin \rho}  (\cos \rho - i \sin \rho) (F + G)
 \nonumber
\\
+  {\epsilon \over \cos \rho} (\cos \rho - i \sin \rho) (F + G)  - M  (F - G)  = 0 \; .
\nonumber
\end{eqnarray}

\noindent
Now summing and subtracting two last  equations, we get
\begin{eqnarray}
 ({d  \over d \rho}    + \nu \;  { \cos \rho \over \sin \rho}   - i \epsilon \; { \sin \rho \over  \cos  \rho}
 ) \; F
+ \;  ( \; \epsilon  + M  - i \nu   - {i \over 2}   )\;  G = 0 \; ,
\nonumber
\\
({d  \over d \rho}    - \nu   \; { \cos \rho \over \sin \rho}
 + i  \epsilon \; { \sin \rho \over  \cos  \rho}  ) \; G
+ ( - \epsilon  + M   + i \nu   - {i \over 2}  )  \; F = 0 \; .
\label{10.17}
\end{eqnarray}

\noindent
This system can be simplified by the substitutions
\begin{eqnarray}
F  (\rho) = \sin ^{-\nu} \rho \; \cos^{-i\epsilon} \rho \;  \varphi  (\rho) \; , \;
G  (\rho) = \sin ^{+\nu} \rho \; \cos^{+i\epsilon} \rho \;  \Gamma   (\rho) \; , \label{10.18}
\nonumber
\\
 \sin ^{-2\nu} \rho  \cos^{-2i\epsilon} \rho   {d \over d \rho} \varphi
+   (  \epsilon  + M  - i \nu   - {i \over 2}   )  \Gamma
= 0 \; ,
\nonumber
\\
 \sin ^{+2\nu} \rho  \cos^{+2i\epsilon} \rho   {d \over d \rho} \Gamma   +
( - \epsilon  + M   + i \nu   - {i \over 2}  )   \varphi   =0 \; .
\label{10.19}
\end{eqnarray}

Let us specify  2nd order differential equations for $\varphi$
\begin{eqnarray}
{d^{2} \over d \rho^{2} } +
 ( -2 \nu \; {\cos \rho \over \sin \rho } +
2i \epsilon\;  {\sin \rho \over  \cos \rho }) {d \over d \rho }
+
[ (  \epsilon  -i \nu )^{2} -( M  - i/2 )^{2}] \;   \varphi   =0 \; .
\nonumber
\end{eqnarray}

\noindent Turning  back to the variable $r = \sin \rho$
\begin{eqnarray}
(1 -r^{2}) {d^{2} \varphi \over dr^{2}} + [\; - {2\nu \over r} + (2\nu -1 + 2i\epsilon) r\; ] \;{d \varphi\over dr }
\nonumber
\\
 +
[ (  \epsilon  -i \nu )^{2} -( M  - i/2 )^{2}] \;   \varphi   =0 \; ,
\nonumber
\end{eqnarray}

\noindent
and transforming the equation  to the variable $z = r^{2}$ one obtains
\begin{eqnarray}
z(1-z) {d^{2}  \varphi \over dz^{2}} + [ ({1 \over 2} - \nu ) - (1 -\nu -i\epsilon ) z  ] {d \varphi \over d z}
\nonumber
\\
-{1 \over 4} [ ( M  - i/2 )^{2}  -(  \epsilon  -i \nu )^{2} ] \;  \varphi = 0 \; .
\label{10.20}
\end{eqnarray}

\noindent
The latter is of a hypergeometric type with parameters defined by
\begin{eqnarray}
\gamma ={1 \over 2} - \nu \; ,
\qquad
 \alpha + \beta  =  -\nu - i\epsilon  \;, \qquad
\alpha \beta = {1 \over 4} [ ( M  - i/2 )^{2}  -(  \epsilon  -i \nu )^{2} ] \ ;
\nonumber
\end{eqnarray}

\noindent
from whence it follows that
\begin{eqnarray}
\alpha = {-\nu - i\epsilon + (iM + 1/2) \over 2} \; , \qquad
\beta = {-\nu - i\epsilon - (iM + 1/2) \over 2} \; .
\nonumber
\end{eqnarray}

Thus, the solutions are given by (remember that $\nu = j +1/2$)
\begin{eqnarray}
\alpha = {-j   - i (\epsilon + M)  \over 2} \; , \qquad
\beta = {-j-1   - i (\epsilon - M)   \over 2} \; ,
\nonumber\\
  \gamma = - j = -1/2, -3/2, ... , \qquad
\varphi (r)  = F(\alpha, \beta, \gamma, z= r^{2}) \; .
\label{10.22}
\end{eqnarray}

The main conclusions about the behavior  of the spin 1/2 particle
in the de Sitter space (near and at far distances from horizon) remain the same as for particles with spin
 $S=0,1$.

\section{Conclusions}

Let us  summarize results.

Exact wave solutions for particles with spin $0, 1/2$ and $1$ in
 the static coordinates of the de Sitter space-time model
 are examined in detail.
For scalar particle, two pairs of linearly independent
solutions are specified explicitly:  running and standing waves.
 A known
 algorithm for calculation of the reflection coefficient
$R_{\epsilon j}$ on the background of the~de Sitter space-time
model is analyzed. It is shown that the
determination of  $R_{\epsilon j}$ requires an additional
constrain on quantum numbers $\epsilon \rho / \hbar c >> j$, where
$\rho$ is a curvature radius. When taken into account of this condition,
the  $R_{\epsilon j}$ vanishes identically.
 It is claimed
that the calculation of the reflection coefficient $R_{\epsilon j}$
is not required at all because there is no barrier in an~effective
potential curve  on the background of the~de Sitter space-time.

The same conclusion holds   for arbitrary particles with higher
spins, it is demonstrated explicitly with the help of exact
solutions for  electromagnetic and  Dirac fields.

\section{Acknowledgement}

Authors are grateful to participants of the  scientific seminar of
the Laboratory of theoretical physics in Institute of physics of   National academy of sciences  of
Belarus for  stimulation discussions and advices.

\end{document}